\documentclass[10pt,twocolumn,twoside,aps,prb,preprintnumbers,reprint]{revtex4-1}
\usepackage{bm}
\usepackage{amsmath}
\usepackage{latexsym,amsfonts}
\usepackage{amssymb}
\usepackage{textcomp}
\usepackage{graphicx}
\usepackage[caption=false]{subfig}
\usepackage[dvipsnames]{xcolor}
\usepackage[colorlinks=true,linktocpage=true,breaklinks=true]{hyperref}
\usepackage{cleveref}

\colorlet{mylinkcolor}{Bittersweet}
\colorlet{mycitecolor}{BurntOrange}
\colorlet{myurlcolor}{MidnightBlue}
\begin{document}
	
\title{Topologically stable bimerons and skyrmions in vanadium dichalcogenide Janus monolayers}
\author{Slimane Laref$^{1*}$}%
\email[]{slimane.laref@kaust.edu.sa}
\author{V. M. L. D. P. Goli$^{1*}$}%
\email[]{durga.goli@kaust.edu.sa}
\author{Idris Smaili$^{2}$}%
\author{Udo Schwingenschl\"ogl$^{1}$}
\author{Aur\'elien Manchon$^{1,2,3}$}
\email[]{manchon@cinam.univ-mrs.fr}
\affiliation{$^1$Physical Science and Engineering Division (PSE), King Abdullah University of Science and Technology (KAUST), Thuwal 23955-6900, Saudi Arabia\\
$^2$Computer, Electrical and Mathematical Science and Engineering (CEMSE), King Abdullah University of Science and Technology (KAUST), Thuwal 23955-6900, Saudi Arabia\\
$^3$Aix-Marseille Univ, CNRS, CINaM, Marseille, France}
	
\begin{abstract} 
We investigate the magnetic phase diagram of 1T-vanadium dichalcogenide monolayers in Janus configuration (VSeTe, VSSe, and VSTe) from first principles. The magnetic exchange, magnetocrystalline anisotropy and Dzyaloshinskii-Moriya interaction (DMI) are computed using density functional theory calculations, while the temperature- and field-dependent magnetic phase diagram is simulated using large-scale atomistic spin modeling in the presence of thermal fluctuations. The boundaries between magnetic ordered phases and paramagnetic phases are determined by cross-analyzing the average topological charge with the magnetic susceptibility and its derivatives. We find that in such Janus monolayers, DMI is large enough to stabilize non-trivial chiral textures. In VSeTe monolayer, an asymmetrical bimeron lattice state is stabilized for in-plane field configuration whereas skyrmion lattice is formed for out-of-plane field configuration. In VSSe monolayer, a skyrmion lattice is stabilized for out-of-plane field configuration. This study demonstrates that non-centrosymmetric van der Waals magnetic monolayers can support topological textures close to room temperature.
\end{abstract}
\maketitle
\section{Introduction}
Magnetic skyrmions, i.e., topologically stable magnetic textures, have recently attracted great interest because of their unique transport and topological properties\cite{nagaosa2013topological,ryu2013chiral,emori2013current}, opening avenues to novel potential applications in the field of spintronics\cite{fert2013skyrmions,Li2017,Huang2017c,Prychynenko2018,Luo2018,Zazvorka2019,Song2020}. Stable skyrmion crystals and metastable isolated skyrmions are normally obtained from the competition between magnetic exchange, uniaxial anisotropy, antisymmetric Dzyaloshinskii-Moriya interaction (DMI) \cite{Moriya1960anisotropic,Dzyaloshinskii1957}, and possibly the external magnetic field. They have been initially reported at low temperature in non-centrosymmetric magnets \cite{Muhlbauer2009,Yu2010} and more recently at room temperature in transition metal multilayers \cite{Jiang2015,Chen2015b,Boulle2016,Moreau-Luchaire2016,Woo2016}. The advent of two-dimensional van der Waals (2D vdW) intrinsic magnets such as Cr$_2$Ge$_2$Te$_6$ \cite{Gong2017}, MnSe$_{2}$\cite{o2018room,O'Hara2018,O'Hara2018b}, VSe$_{2}$\cite{bonilla2018strong}, CrI$_{3}$\cite{huang2017layer}, and Fe$_{3}$GeTe$_{2}$\cite{fei2018two} opens appealing avenues for the observation of chiral magnetic textures in atomically thin materials. What makes 2D vdW magnets particularly attractive is the possibility to engineer their band structure by surface chemistry. An outstanding demonstration of this feature is the synthesis of transition metal Janus monolayers \cite{Cheng2013b,lu2017janus,Zhang2017}, where the transition metal element is embedded between dissimilar (chalcogen or halide) ions (see also Ref. \onlinecite{Zhang2020c}). This configuration breaks the inversion symmetry and promotes Rashba-type spin-orbit coupling \cite{Cheng2013b,Din2019,Chen2020}. In the case of {\em magnetic} Janus monolayers, the inversion symmetry breaking enables spin-orbit torque \cite{Manchon2019,Smaili2020} as well as DMI \cite{Liang2020,Zhang2020d,Yuan2020c}.

In the present work, we investigate the onset of DMI and the emergence of magnetic skyrmions in vanadium-based transition metal dichalcogenide Janus monolayers in 1T configuration. Specifically, we focus our investigation in VSSe, VSTe and VSeTe. In Section \ref{s:dmi}, we study the magnetic exchange interaction and anisotropy from first principles and compute DMI using the generalized Bloch theorem. In Section \ref{s:atomistics}, we investigate the magnetic phase diagram of these systems under the combined action of an external magnetic field and thermal excitations using an atomistic spin simulation method. By exploring in details the temperature-field magnetic phase diagram of the three systems, we find that chiral ground states can be obtained in a reasonably large range of temperatures. In particular, VSeTe displays asymmetrical bimeron lattice ground state up to room temperature, whereas VSSe exhibits skyrmion lattice states.

\begin{figure}
\centering
		\includegraphics[width=\linewidth]{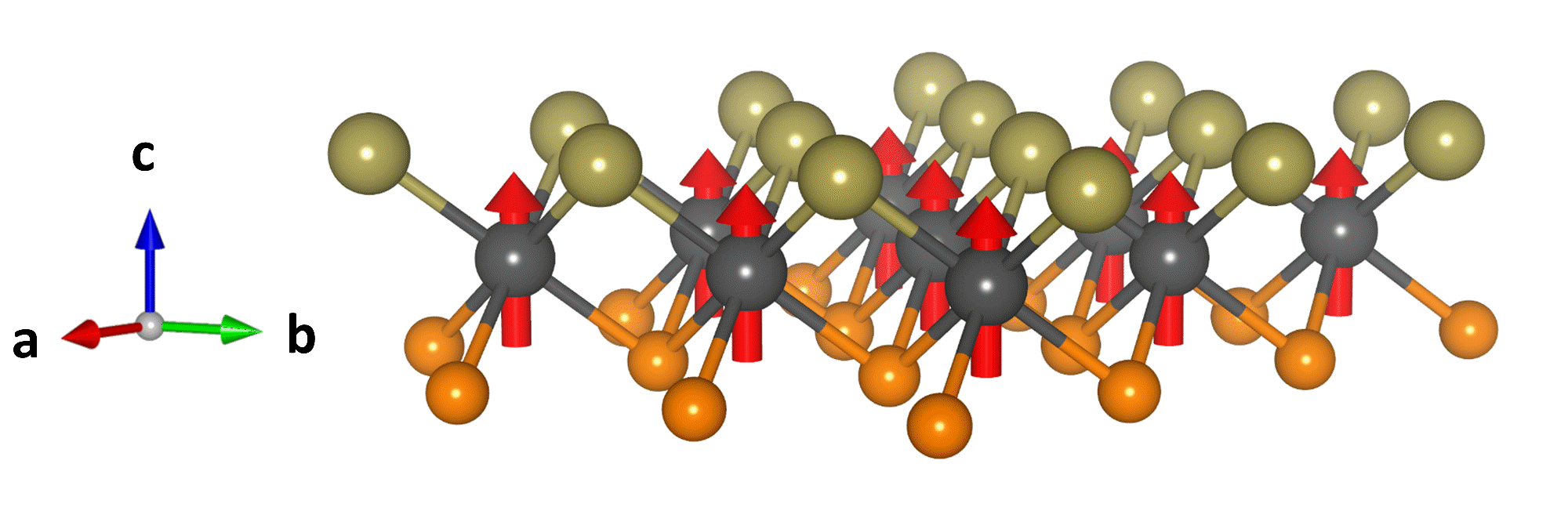} 
		\caption{(Color Online) Cartoon of the VXY Janus monolayer. Vanadium elements are represented in grey, and the chalcogen elements X and Y (S, Se, and Te (X$\neq$Y)) are in orange and brown, respectively. The red arrows indicate the direction of the magnetic moment.\label{fig1}}
\end{figure}

\section{First principles calculations\label{s:dmi}} 
\subsection{Structural and magnetic properties} 
To compute the structural and magnetic properties, we perform first-principles calculations using the full-potential linearized augmented plane-wave (FLAPW) method as implemented in FLEUR code\cite{hamann1979semiconductor,wimmer1981full,flapw}. Our calculations are performed using the local-density approximation exchange-correlation functional (LDA-vwn) as implemented in the FLAPW Package\cite{flapw}. For the angular momentum expansion and the reciprocal plane wave, cut-off of $I_{max}$ = 6 and $k_{max}$ = 4 were applied, and we used $radii$ of MT spheres around 1.7 a.u for S, 2.1 a.u for Se and 2.4 a.u for Te, and 2.2 a.u for V, where a.u is the Bohr $radius$. ${\Gamma}$-centered k-grid $32\times32\times1$ and $64\times64\times1$ have been sampled for the whole Brillouin zone to achieve the total energy without spin-orbit coupling (SOC) and with SOC, respectively.\par

By defining ground-state geometries, crystal structures of the bulk VXY (X, Y= S, Se, Te) have been optimized. Table \ref{tab1} lists the structural and magnetic parameters of VSeTe, VSSe, and VSTe materials, which are in good agreement with the literature \cite{Ma2012evidence,Feng2012giant,Zhang2013dimension,Pan2014electronic}. All three Janus materials exhibit C$_{3v}$ symmetry and the on-site magnetic moments of V is controlled by the electron depletion due to the ionic bonding with the chalcogen elements. As a phenomenological rule, the larger the electronegativity difference between X and Y elements, the larger the charge depletion, and the larger the magnetic moment. The magnetic exchange $J$ is calculated from the energy difference between the ferromagnetic and antiferromagnetic state. We find that all three systems are ferromagnetic and $J$ roughly scales with the electronegativity difference $\Delta\chi$, yielding the largest value (70 meV) for VSTe. Magnetocrystalline anisotropy is obtained by utilizing the force theorem and applying SOC within the second variation method \cite{oswald1985interaction,liechtenstein1987local,li1990magnetic}. The magnetocrystalline anisotropy, $K=E_{\parallel}-E_{\perp}$, defined as the difference between in-plane and out-of-plane magnetization energies is reported in Table \ref{tab1}. Our results indicate that VSSe possesses a weak out-of-plane uniaxial anisotropy, whereas both VSeTe and VSTe exhibit {\em easy-plane} anisotropy (in other words, there is no variation of the magnetic energy when rotating the magnetization in the ($x,y$) plane for these two monolayers).

	\begin{table*}
		\centering
		\caption{Structural and magnetic parameters of T-VXY: The lattice constant (a), the distance between X and Y ($d_{\rm X-Y}$), the electronegativity difference between the two chalcogen elements ($\Delta\chi$ - defined positive), the magnetic moment of the transition metal atom ($\mu_s$), the  Heisenberg exchange ($J$), the magnetocrystalline anisotropy ($K$), and DMI strength ($D$). The Curie temperature ($T_{\rm c}$) is deduced from zero-field susceptibility analysis as explained in Section \ref{s:atomistics}. \label{tab1}} \
		\begin{tabular}{c c c c c c c c c} \hline 
			{VXY} & {a}({\AA}) & {$d_{\rm X-Y}$}({\AA}) & $\Delta\chi$ &  {$\mu_s$}($\mu_B$) & $J$(meV) & $K$(meV) & $D$ (meV$\cdot${\AA})& $T_{\rm c}$(K) \\ \hline 
			{VSSe}   & 3.266 & 2.969 & 0.03 & 0.686 &20.3  & 0.078  & 1.9  & 240  \\
			{VSeTe}  & 3.673 & 2.887 & 0.45 & 1.391 & 50.82 & -0.963 & -5.7 & 630 \\
			{VSTe}   & 3.611 & 2.838 & 0.48 & 1.405 & 71.51 & -0.860 & 2.5  & 780 \\  
			 \hline 
		\end{tabular}
	\end{table*}
	
\subsection{Spin spiral calculations}

\begin{figure}
\centering
		\includegraphics[width=\linewidth]{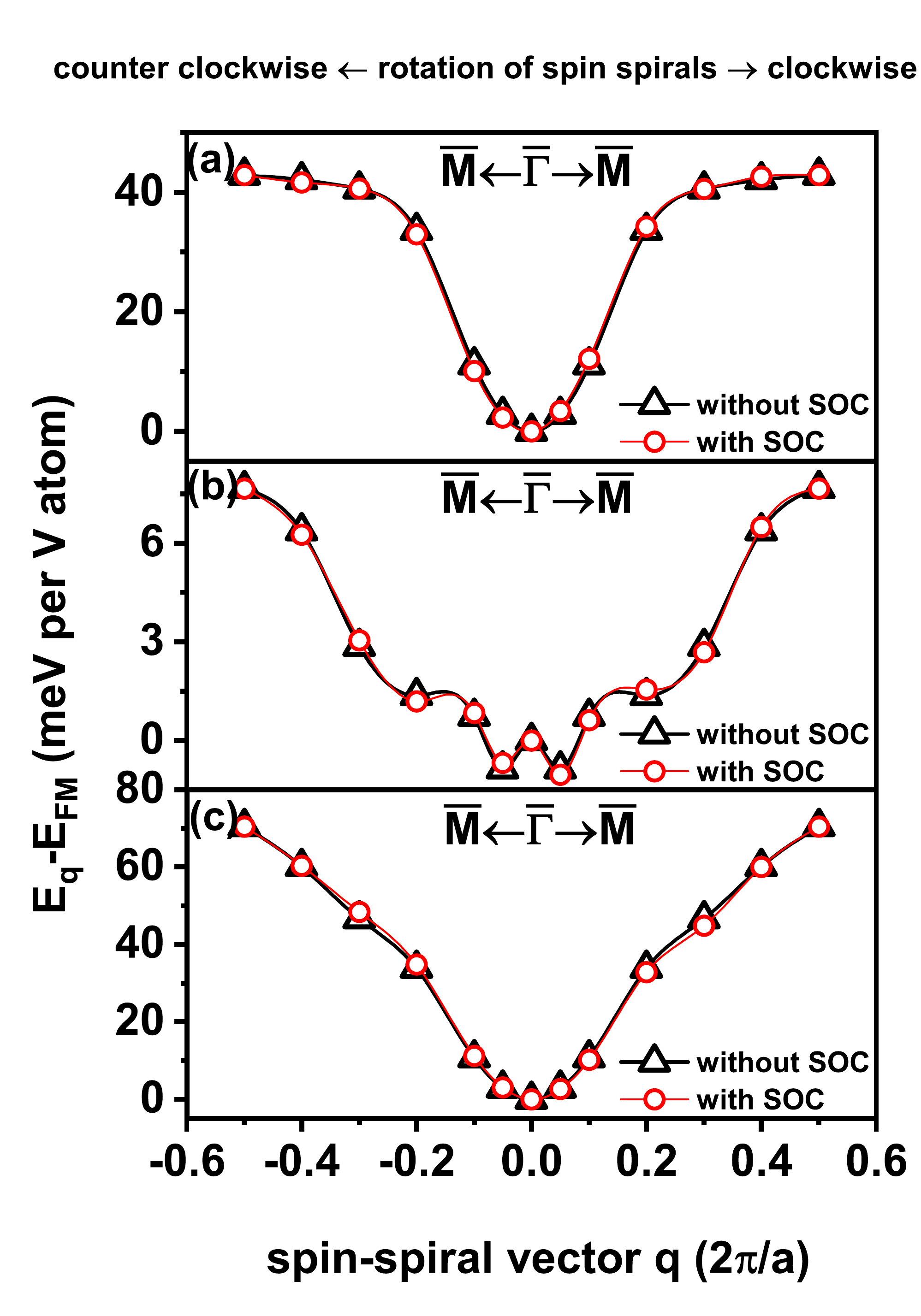}
		\caption{(Color Online) Calculated energy dispersions $E_{q}$ of rotating spin spirals without spin-orbit coupling (black symbols) and with
			spin-orbit coupling (red symbols) of (a) VSeTe (b) VSSe, and (c) VSTe. \label{fig3}}
\end{figure} 

\begin{figure}
\centering
		\includegraphics[width=\linewidth]{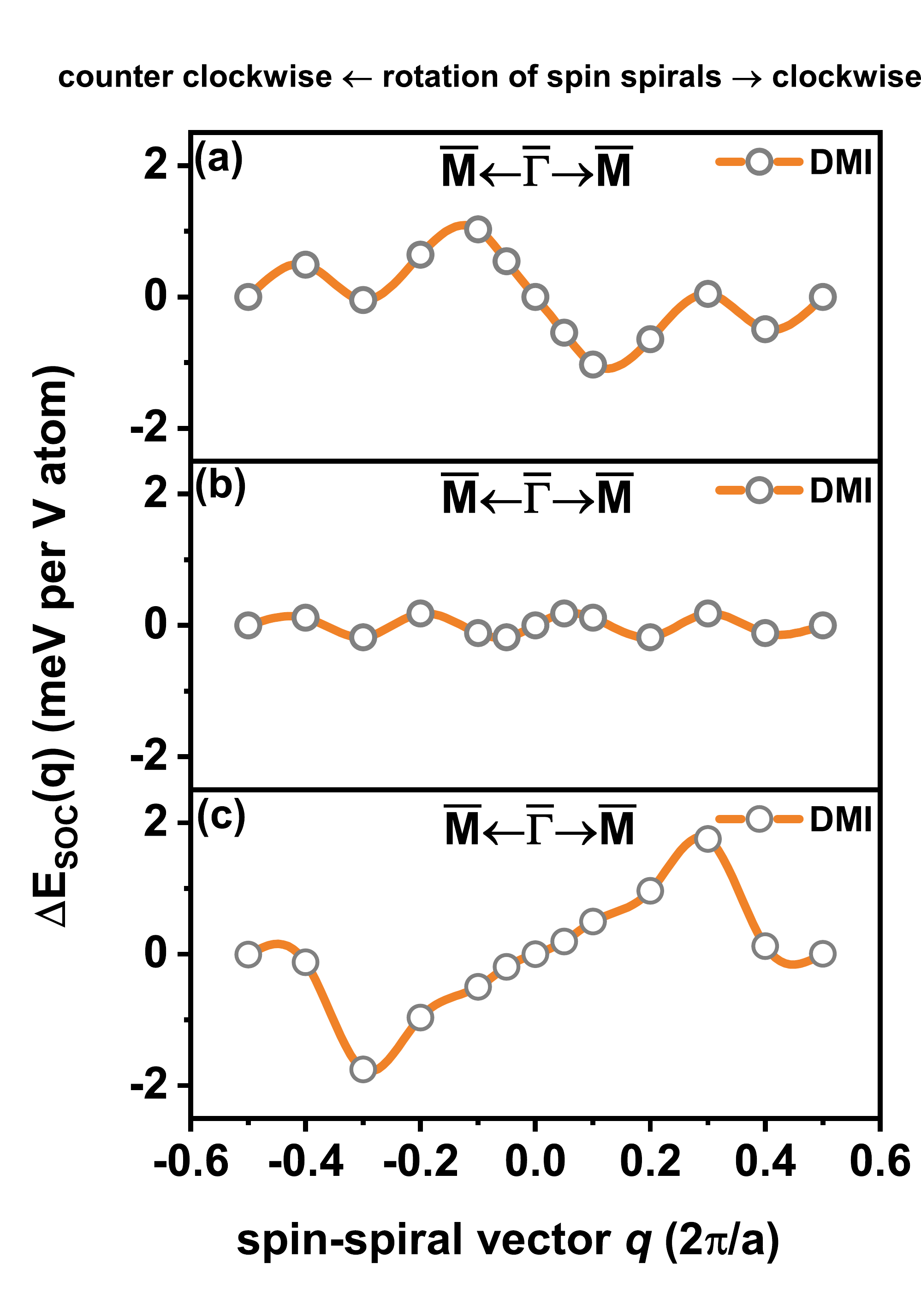}
		\caption{(Color Online) Antisymmetric contribution to the energy dispersion displayed on Fig. \ref{fig3} as a function of the spin-spirals vector $q$ for (a) VSeTe (b) VSSe, and (c) VSTe.\label{fig4}}
\end{figure}

DMI is an antisymmetric exchange interaction that tends to align neighboring magnetic moments perpendicular to each other. It reads
\begin{align}
{E}_{\rm DM}=\sum_{ij}{\bf D}_{ij}\cdot\left( {\bf S}_{i} \times {\bf S}_{j} \right), \label{eq1}
\end{align}
where ${\bf S}_{i}$ is the magnetic moment on site $i$, and ${\bf D}_{ij}$ is the Dzyaloshinskii-Moriya vector that governs the DMI between sites $i$ and $j$. Because DMI is controlled by inversion symmetry breaking, it is {\em odd} in SOC strength and therefore, a good estimation of ${\bf D}_{ij}$ is obtained at the first order in SOC. To do so, we exploit the generalized Bloch theorem \cite{sandratskii1991symmetry} approach implemented in FLEUR code\cite{Kurz2004,Heide2009,zimmermann2014first}. We build out-of-plane N\'eel spin spirals in momentum space whose energy dispersion, without SOC (black) and at the first order in SOC (red), is reported on Fig. \ref{fig3}. The corresponding difference between the spin spiral energy dispersion with and without SOC is reported on Fig. \ref{fig4} and the Dzyaloshinskii-Moriya vector is estimated by taking the slope of the dispersion at $q=0$. This value corresponds to the DMI strength experienced by magnetic textures whose length scale is much larger than the crystal lattice parameter. The extracted values are -5.7 meV$\cdot${\AA}, 1.9 meV$\cdot${\AA} and 2.5 meV$\cdot${\AA} for VSeTe, VSSe and VSTe, respectively. For the sake of comparison, the DMI obtained for Pt/Co(111) using the same method \cite{Belabbes2016} is about 50 meV$\cdot$\AA~due to the large SOC strength of Pt. As discussed in the previous section, the DMI in vanadium-based Janus monolayers is sufficiently strong to stabilize chiral magnetic textures.


\section{Magnetic phase diagram from atomistic spin dynamics\label{s:atomistics}}
\subsection{Methodology} 

To model the magnetic phases of all three VXY systems, we consider atomic spins on each site of a triangular lattice. 
The Heisenberg spin Hamiltonian is given by
\begin{align} 
\mathcal{H}~=~& -J \sum_{\langle i,j\rangle}{\bf S}_i \cdot {\bf S}_j - \sum_{\langle i,j\rangle} {\bf D}_{ij} \cdot\left( {\bf S}_{i} \times {\bf S}_{j} \right)\nonumber \\ 
 & - K\sum_{i}({\bf S}_i \cdot \hat{z})^2 - \mu_s \sum_i {\bf S}_i \cdot {\bf B},
\label{atomistic_hamiltonian}
\end{align} 
\noindent 
where ${\bf S}_i$ is normalized unit spin vector at site $i$. On the right side of Eq. \eqref{atomistic_hamiltonian}, the first term is the Heisenberg exchange energy with ferromagnetic exchange strength $J$ and $\langle i,j\rangle$ indicates the sum over all nearest-neighbors. The second term is the DMI, ${\bf D}_{ij} = D({\bf r}_{ij} \times \hat{z})$, with $D$ is the DMI energy and ${\bf r}_{ij}$ is the unit vector between sites $i$ and $j$. The third term represents the magnetocrystalline anisotropy energy that can be either easy-plane ($K<0$ for VSeTe and VSTe) or uniaxial out-of-plane ($K>0$ for VSSe). The last term is the Zeeman interaction energy due to the applied field, ${\bf B}$. \par
\begin{figure*}[t]
\begin{center} \includegraphics[width=0.9\textwidth]{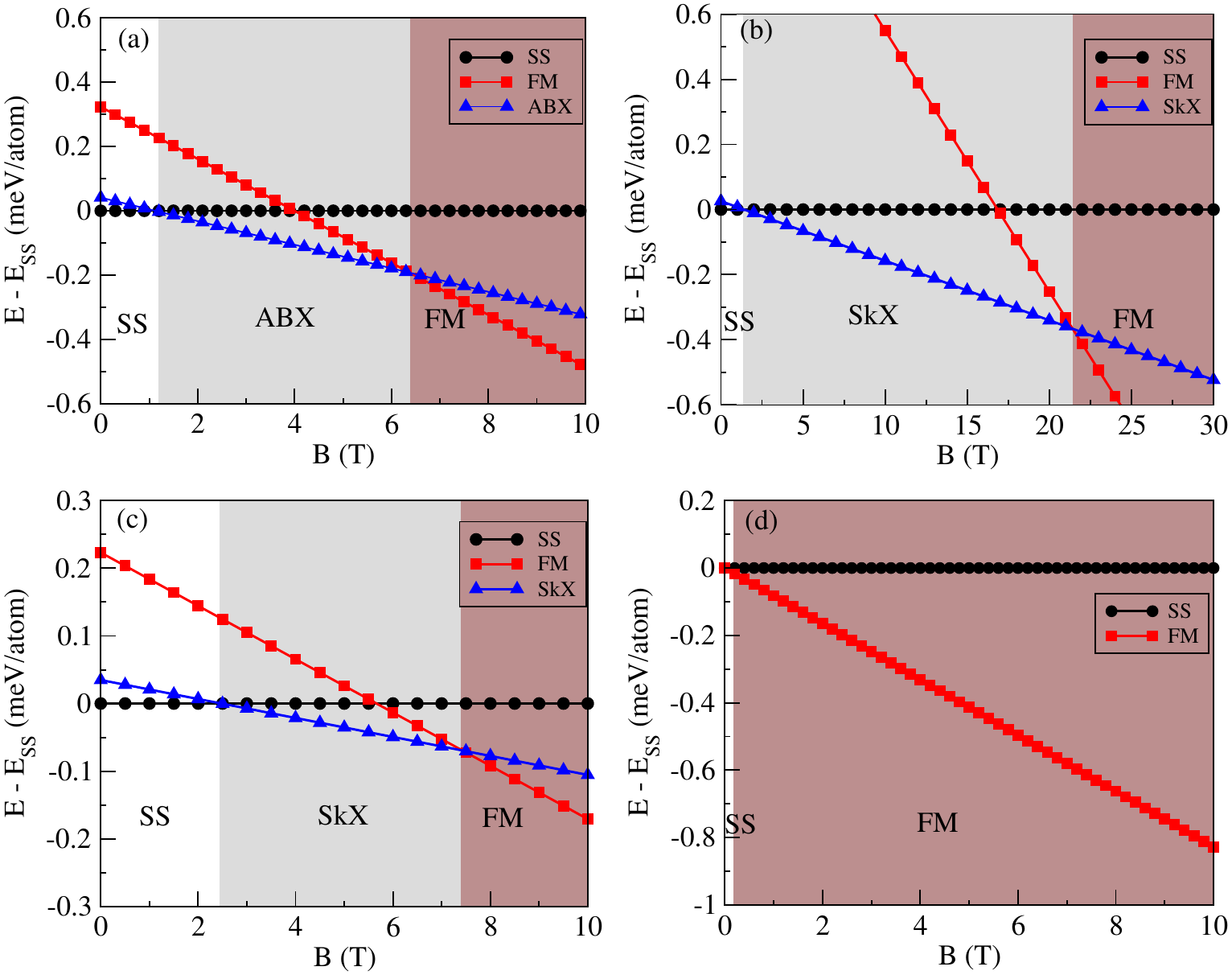}
\caption{Zero temperature phase diagrams of VSeTe, VSSe and VSTe systems. 
The energies corresponding to FM, ABX and SkX states are shown with respect to the relative energy of SS state as a function of the applied magnetic field. 
(a) and (b) represent the phase diagrams of VSeTe with in-plane and out-of-plane applied magnetic field, respectively. The phase diagram of VSSe with out-of-plane magnetic field is shown in (c) while the phase diagram of VSTe is shown in (d) where in-plane magnetic field is considered. The shaded areas represent different ground states.}
\label{zerotemp_phasediagram}
\end{center} \end{figure*}
To obtain the magnetic phase diagrams, we use atomistic spin dynamics technique that numerically solves the stochastic Landau-Lifshitz-Gilbert (LLG) equation of the form
\begin{align}
\frac{\partial {\bf S}_i}{\partial t}=&-\frac{\gamma}{1+\alpha^2} {\bf S}_i \times \left({\bf H}_i^{\rm eff} + \alpha~{\bf S}_i \times {\bf H}_i^{\rm eff}\right),
\label{llgeqn}
\end{align}
\noindent where $\gamma$ is gyromagnetic ratio and $\alpha$ is the intrinsic Gilbert damping constant. 
We choose $\alpha$ = 0.05. The effective field acting on the magnetic moment ${\bf S}_i$ is represented by ${\bf H}_i^{\rm eff}=-\frac{1}{\mu_s}\frac{\partial \mathcal{H}}{\partial {\bf S}_i} + {\bf H}_i^{\rm th}$, where $\mu_s$ is the atomic magnetic moment. ${\bf H}_i^{\rm th}$ is the stochastic thermal field arising from thermal fluctuations of the magnetic moments. By using the Langevin dynamics approach, random thermal field is added at each site. This field obeys the Gaussian distribution in three dimensions with mean of zero, ${\boldsymbol \Gamma}(t)$. The random thermal field is given by \cite{evans2014} 
\begin{align}
{\bf H}_i^{\rm th} = {\boldsymbol \Gamma}(t) \sqrt{\frac{2\alpha K_B T}{\gamma \mu_s \Delta t}},
\end{align}
where $k_B$ is the Boltzmann constant, $T$ is the temperature and $\Delta t$ is integration time step. The simulations are carried out on a triangular lattice with $N = 150\times150$ sites and periodic boundary conditions are implemented. The magnetic parameters are taken from Table \ref{tab1}. In order to characterize the magnetic state at a given point in the ($T,B$) phase space, Eq. \eqref{llgeqn} is solved numerically to obtain the magnetic susceptibility and average topological charge for VXY systems. The magnetic susceptibility is given by 
\begin{align}
\chi = \frac{\mu_s}{k_B T} \left(\langle M^2 \rangle - \langle M \rangle^2\right),
\label{susceptibility_eqn}
\end{align}
where $M = \frac{1}{N} \sum_i {\bf S}_i$. A simple magnetic ground state such as ferromagnetic or antiferromagnetic state and a skyrmion state can be differentiated 
by its topological charge ($Q$). In the continuum magnetization limit, $Q$ is defined by the following equation
\begin{align}
 Q = \frac{1}{4\pi} \int {\bf S} \cdot \left(\frac{\partial {\bf S}}{\partial x} \times \frac{\partial {\bf S}}{\partial y}\right) dxdy.
\end{align}
Here $Q$ describes the number of times magnetic moments (${\bf S}$) wrap around the unit sphere. 
$Q=0$ denotes a trivial state while $Q\neq0$ provides the number of skyrmions present in the skyrmion state.
In a magnetic state with topologically protected magnetic texture analogous to skyrmion, $Q$ is non-zero.
For the discrete square lattice, Berg and L\"uscher proposed a method to quantify $Q$ \cite{berg1981}.
This method involves partitioning the lattice into nearest-neighbor triangles with spins at vertices of each triangle. The same procedure can be applied for triangular lattice except that partitioning the lattice into triangles is not required. Counter-clockwise rotation of vertices spins ${\bf S}_i$, ${\bf S}_j$ and ${\bf S}_k$ of each triangle is considered to calculate 
$Q$ from the following equation \cite{bottcher2018,muller2019},
\begin{align}
\mathrm{tan}\left(Q/2\right) = \frac{{\bf S}_i \cdot ({\bf S}_j \times {\bf S}_k)} {1+{\bf S}_i \cdot {\bf S}_j + {\bf S}_j \cdot {\bf S}_k + {\bf S}_k \cdot {\bf S}_i}.
\label{tq_eqn}
\end{align}

The spin dynamics simulations are performed with increasing temperature for a fixed magnetic field. 
At each point in ($T,B$) phase space, the system is allowed to evolve for 10$^6$ time steps, 
and the average topological charge and magnetic susceptibility are calculated for 3$\times$10$^6$ averaging time steps. 
The considered simulation time is sufficient to capture all variations of temperature-dependent quantities.

\subsection{Zero-temperature phase diagram}

We first compute the zero-temperature ground states of VSeTe, VSSe and VSTe systems under an external applied field. Depending on the field strength, we find that various types of magnetic phases can be stabilized: ferromagnetic (FM), spin spiral (SS), skyrmion crystal (SkX) as well as asymmetric bimeron crystal (ABX). The field-dependence of these various phases at zero temperature is reported on Fig. \ref{zerotemp_phasediagram}. 
In this figure, we report the results for VSeTe (easy-plane anisotropy) with both in-plane (a) and out-of-plane magnetic fields (b), VSSe (out-of-plane uniaxial anisotropy) with out-of-plane magnetic field (c), and VSTe (easy-plane anisotropy) with in-plane magnetic field (d).

\begin{figure}[htbp]
\begin{center} \includegraphics[width=08.8cm,height=10.0cm]{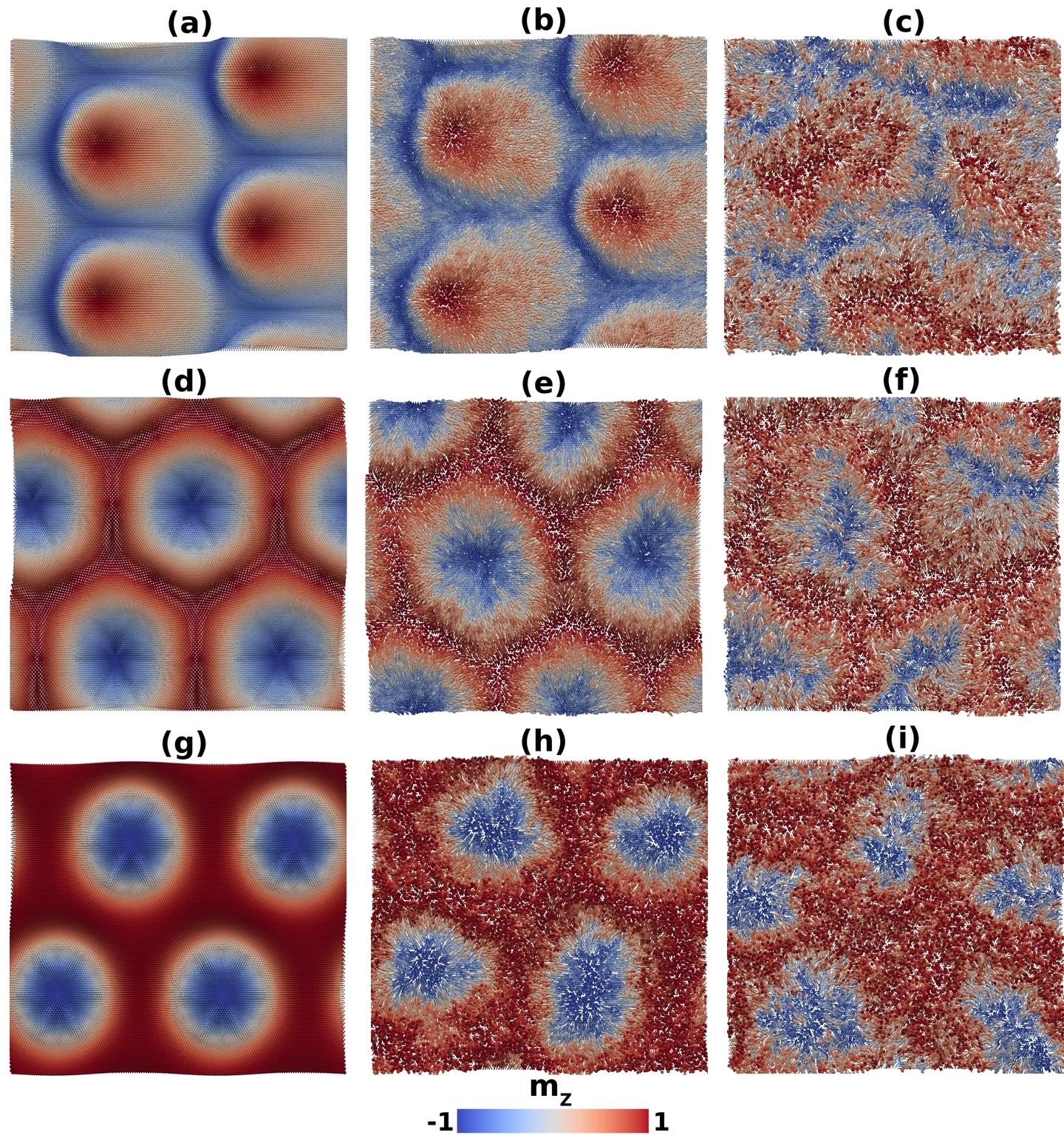}
\caption{Spin textures of VSeTe with in-plane and out-of-plane fields and VSSe at different temperature and field combinations.
At in-plane field 2 T,  ABX state of VSeTe with temperatures (a) T = 0 K, (b) T = 60 K, and (c) T = 300 K.
At out-of-field 3 T, SkX state of VSeTe with temperatures (d) T = 0 K, (e) T = 60 K, and (f) T = 300 K.
SkX state of VSSe with temperatures (g) T = 0 K, (h) T = 60 K and (i) T = 150 K at B = 3 T.}
\label{vxy_spin_textures}
\end{center} \end{figure}

In the case of VSeTe, we find SS and FM states in the low field and high field limits, respectively, independently on the field direction. Interestingly, at intermediate field we obtain two topologically non-trivial lattices, namely asymmetrical bimeron lattice [ABX - Fig. \ref{zerotemp_phasediagram}(a)] for in-plane field configuration and skyrmion lattice [SkX - Fig. \ref{zerotemp_phasediagram}(b)] for out-of-plane field configuration. Asymmetrical bimerons are planar counterpart of magnetic skyrmions that are known to emerge under the cooperation of DMI with uniaxial in-plane anisotropy \cite{lshen2020}. In VSeTe, which possesses easy-plane rather than uniaxial in-plane anisotropy, the in-plane magnetic field is necessary to stabilize the bimeron lattice. Notice that the magnetization of the SS state is nearly vanishing and hence this state is mostly unaffected by the magnetic field as shown in Fig \ref{zerotemp_phasediagram}. In contrast, both ABX and FM states have non-zero magnetization and their energy decreases with increasing field. In our 150$\times150$ spins system, the ABX state with four bimerons becomes the ground state for fields above 1.2 T. The spin texture of ABX state is shown in Fig \ref{vxy_spin_textures}(a). It is also possible to promote lattices with more bimerons with different sizes; however, the energy of these bimeron states remain higher than ABX state with four bimerons and are therefore metastable. Therefore, the energy of ABX state depends on the number of bimerons and on their size. With increasing field, the size of bimerons shrinks. However, ABX state maintains lower energy until FM state becomes the ground state. The magnetization of FM state is larger than that of ABX state, and hence the energy of FM state lowers swiftly for large magnetic fields. 
Above 6.4 T, VSeTe acquires in-plane magnetized state. In the case of out-of-plane magnetic field configuration of VSeTe, SS state is the lowest energy state below 1.3 T as shown in Fig. \ref{zerotemp_phasediagram}(b). The SkX state becomes the ground state over a very large range of applied field, between 1.3 T and 21.4 T. Noticeably, the skyrmions' shape is hexagonal in this range, due to large skyrmion-skyrmion interaction [see Fig \ref{vxy_spin_textures}(d)]. In the SkX ground state, the skyrmion diameter reduces from 21 nm to 12 nm upon increasing field. The domain wall width of skyrmion is large due to easy-plane anisotropy. Above 21.4T though, all spins of VSeTe align to produce the FM ground state. 

In the case of VSSe, shown in Fig. \ref{zerotemp_phasediagram}(c), a SkX state with four skyrmions constitutes the ground state between 2.5 T and 7.4 T. In this state, the skyrmion diameter changes from 18 nm to 11 nm upon increasing field. Below 2.5 T, the SS state is the ground state while the magnetization is saturated for fields above 7.4 T. In the case of VSTe, shown in Fig. \ref{zerotemp_phasediagram}(d), the SS state is the ground state in the absence of in-plane field. Due to the presence of easy-plane anisotropy and small DMI, VSTe has in-plane spin texture with out-of-plane tilting. However, the energy gap between SS and FM states is fairly small. VSTe acquires fully magnetized state with the application of small in-plane field. 

\begin{figure*}[htbp]
\begin{center} 
\includegraphics[width=\textwidth]{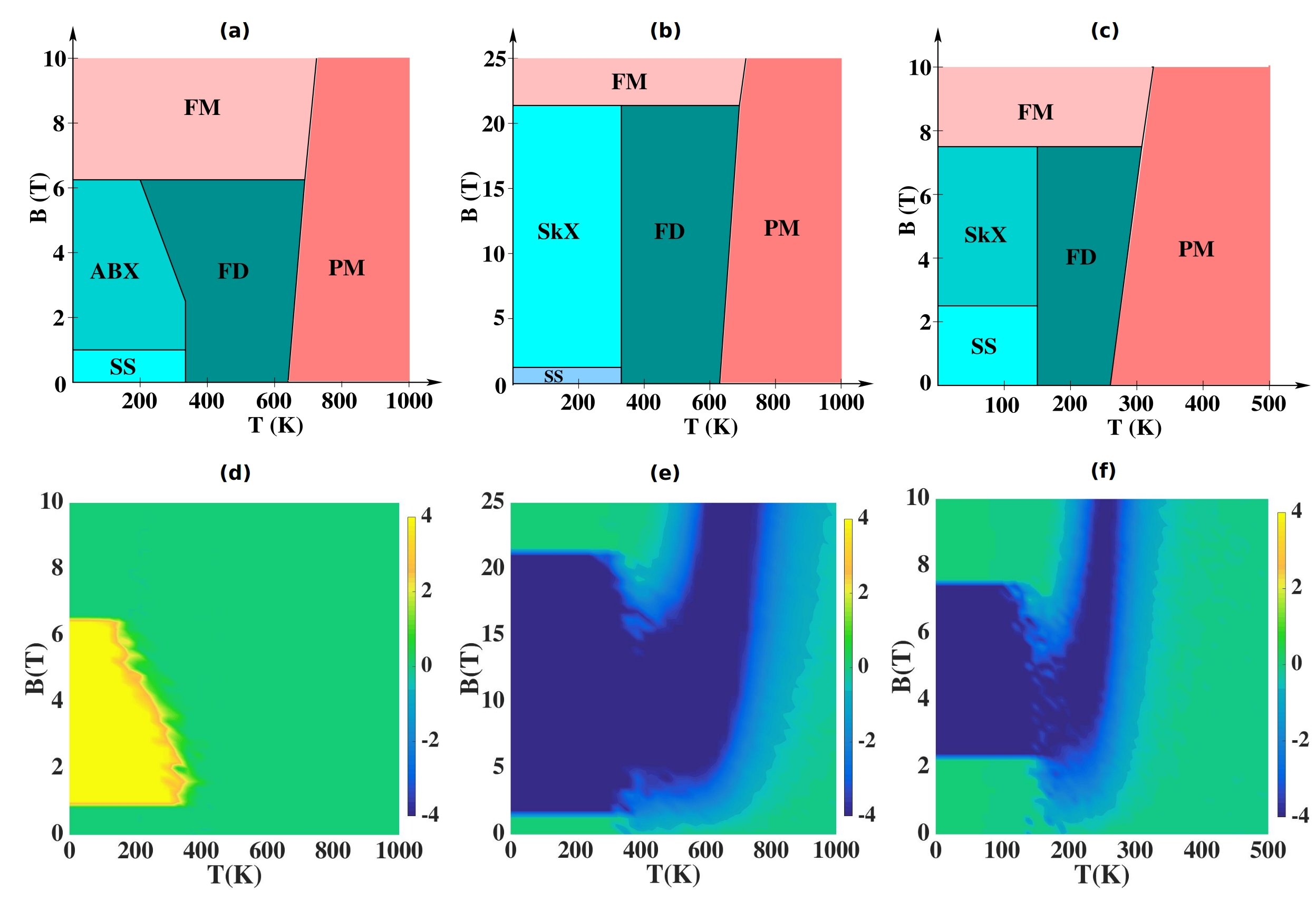}\\
\caption{(Top) ($T,B$) phase diagram and (Bottom) corresponding topological charge $Q$ for VSeTe with (a,d) in-plane and (b,e) out-of-plane fields and (c,f) VSSe systems.}
\label{vxy_phasediagram}
\end{center} \end{figure*}

\subsection{($T,B$) magnetic phase diagram}

The simulations are performed up to a maximum temperature of 1500 K and a maximum field of 25 T. The average topological charge and magnetic susceptibility are calculated to define the phase boundaries and critical temperatures. The phase diagrams of the three systems and their corresponding average topological charge are shown in Fig. \ref{vxy_phasediagram}. Along with ABX or SkX, two new phases emerge: fluctuation-disorder (FD) and paramagnetic (PM) phases. 
In the FD region, both bimerons and skyrmions acquire a finite lifetime and hence the average topological charge remains non-zero \cite{rozsa2016}. In contrast, the system goes into magnetic disordered state in the PM region and exhibits zero topological charge. The details of these two phases are discussed below.
The boundaries between all phases are determined by the inflection points of the temperature-dependent magnetic susceptibility and its derivatives, as discussed by Ref. \onlinecite{grigoriev2011}.

\subsubsection{VSeTe Phase Diagram}

In the case of VSeTe with in-plane field configuration, displayed on Fig. \ref{vxy_phasediagram}(a), the system is stabilized in cycloidal SS state at low temperature and zero magnetic field. In this state, the magnetic domains are connected by N\'eel-type domain walls. As the in-plane field approaches 1.2 T, all the domains convert into ABX state. In Fig. \ref{vxy_spin_textures}(a), the magnetic texture of ABX state is shown with B = 2 T and T = 0 K. As the magnetic field increases to 6.4 T, the ordered ABX state breaks down into FM ground state. Above this field, bimeron states can be found. However, from the zero temperature phase diagram of VSeTe in Fig \ref{zerotemp_phasediagram}(a), these states are considered as metastable ABX states and they remain higher energy states. Hence, for fields above 6.4 T, all bimerons dissolve and the FM state becomes the ground state.

Because a bimeron is a topologically protected magnetic texture analogous to skyrmion, it can be identified by a non-zero topological charge ($Q$). A bimeron with charge $Q$ can be converted into another bimeron of opposite sign, $-Q$, merely by changing the sign of its magnetic and spatial components. Indeed, a bimeron with charge $Q$ identified by its spatially dependent magnetic configuration, [S$_x$(x,y),S$_y$(x,y),S$_z$(x,y)], can be converted into another bimeron of charge $-Q$ with magnetic configuration [S$_x$(-x,y),-S$_y$(-x,y),-S$_z$(-x,y)] \cite{lshen2020}. In addition, the coexistence of two bimerons with opposite signs of $Q$ can occur without loss of state energy. Using Eq. \eqref{tq_eqn}, the average charge $Q$ is calculated in the ($T,B$) phase diagram and is shown in Fig. \ref{vxy_phasediagram}(d). The finite value of average $Q$ represents the ABX region and its boundary with the other topologically-trivial regions. We obtain $Q= 4$ which indicates that four bimerons present in the ABX state, a number that remains uniform throughout the ABX region at low temperatures, as shown in Fig. \ref{vxy_phasediagram}(a).

\begin{figure}[htbp]
\begin{center} \includegraphics[width=09.0cm,height=7.0cm]{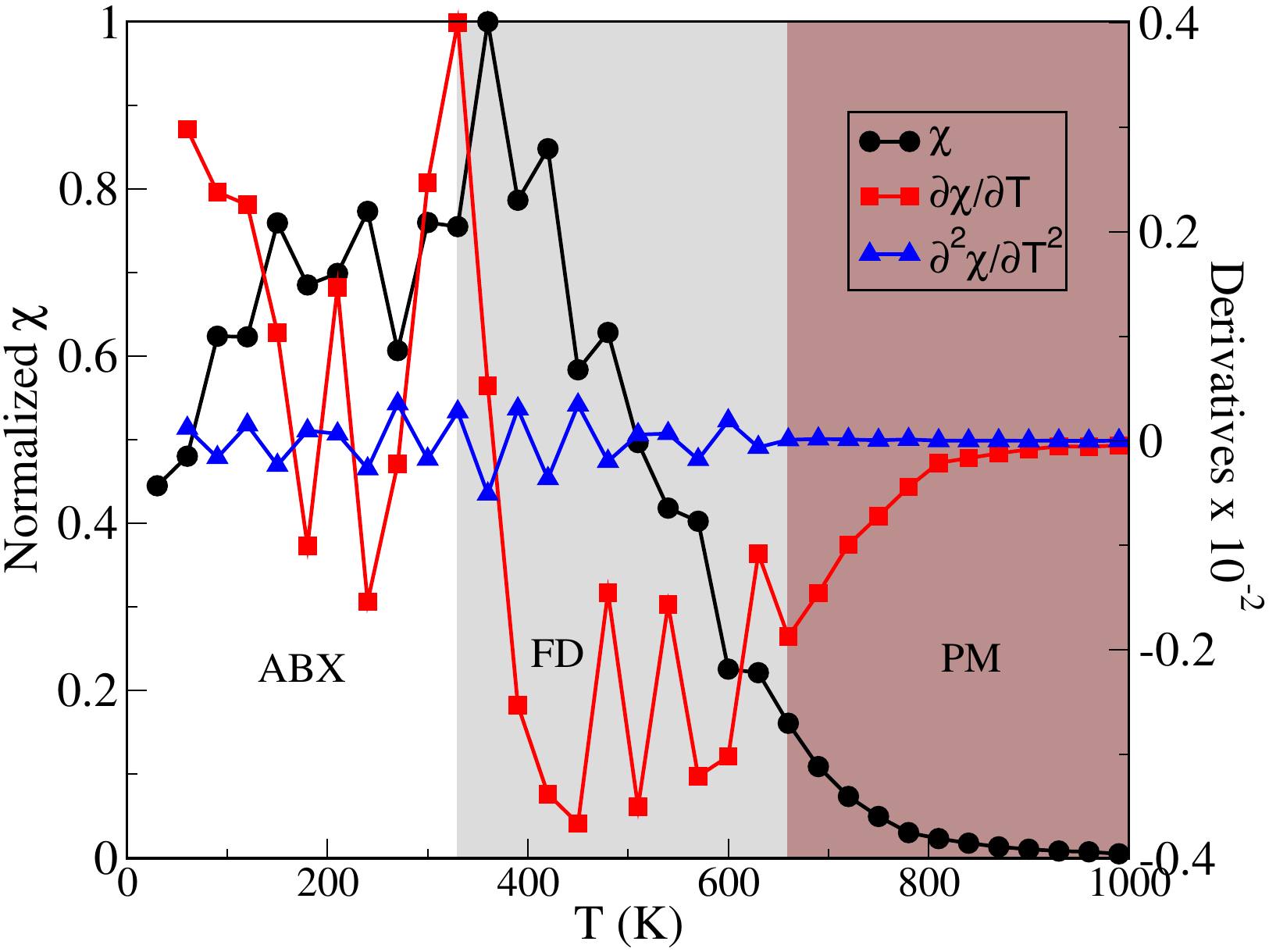}
\caption{
Temperature dependent magnetic susceptibility and its first and second order derivatives of VSeTe system with in-plane field 2.6 T.
The infection points of $\frac{\partial \chi}{\partial T}$ separate three phases, namely, ABX, FD and PM phases respectively.}
\label{susceptibility_diagram}
\end{center} \end{figure}

In order to investigate temperature dependent phases, we plot the normalized magnetic susceptibility and its first and second derivatives as a function of
temperature at specific applied fields. Figure \ref{susceptibility_diagram} shows the susceptibility and its derivatives as a function of temperature at B = 2.6 T. It is difficult to determine the phase boundaries using the normalized susceptibility curve. However, $\frac{\partial \chi}{\partial T}$ displays a jump at the phase transition. The inflection points of $\frac{\partial \chi}{\partial T}$ separate three regions, i.e., ABX, FD and PM phases. The temperature corresponding to the maximum and minimum values of $\frac{\partial \chi}{\partial T}$ determine the boundaries between the phases. The FD region comes after the ABX phase upon increasing temperature. The maximum value of $\frac{\partial \chi}{\partial T}$ shows the lower limit of the FD region at 330 K at lower fields. This transition temperature is field-dependent. This region is known to display skyrmions with finite lifetime \cite{rozsa2016}. This can also be applicable to bimerons as well. In this region, because spins are excited by thermal fluctuations, creation and annihilation of bimerons of opposite signs of $Q$ occurs, resulting in a fluctuation of $Q$ between -4 to +4 during the simulation. On average, the coexistence of bimerons of opposite topological charge leads to $Q\approx0$, as shown in Fig. \ref{vxy_phasediagram}(d). The minimum value of $\frac{\partial \chi}{\partial T}$ in Fig. \ref{susceptibility_diagram} before the convergence represents the Curie temperature of VSeTe and it is 660 K at 2.6 T field. At zero applied field, the Curie temperature of VSeTe is 630 K.

In the case of out-of-plane applied field, the topological charge of VSeTe as a function of temperature and applied field is shown in Fig. \ref{vxy_phasediagram}(b).
At low temperature and in SkX region, we obtain $Q$ = -4, i.e., four skyrmions in SkX state. The FD region starts at 330 K and is almost independent of the applied field.
In this region, the average $Q$ remains constant in contrast to the ABX state. This observation means that the number of skyrmions remains constant and no skyrmions with opposite charge $-Q$ are created in the FD region. The critical temperature of VSeTe is 630 K in the out-of-plane field configuration. Therefore, at low temperatures the critical temperature of VSeTe is independent of the applied field direction.

\subsubsection{VSSe and VSTe Phase Diagrams}

The magnetic phase diagram of VSSe for an external field applied out of the plane is shown in Fig \ref{vxy_phasediagram}(b). The shape of the skyrmions remains circular due to out-of-plane anisotropy [see Fig \ref{vxy_spin_textures}(g)]. From Fig. \ref{vxy_phasediagram}(e), $Q$ = -4 which indicates that four skyrmions are present in the ground state of SkX state.
From the average $Q$ and $\frac{\partial \chi}{\partial T}$ calculations, the FD region starts at 150 K. Although creation and annihilation of skyrmions occur in this region, the average $Q$ remains constant which suggests that only skyrmions with the same sign of $Q$ are present at any time. The inflection point of $\frac{\partial \chi}{\partial T}$ has a minimum at 240 K, which is the critical temperature at zero field.

Let us finally comment on VSTe, whose phase diagram is reported on Fig. \ref{vxy_phasediagram}(c). This system also possesses easy-plane anisotropy but has a much larger exchange than VSeTe. For this reason, VSTe becomes ferromagnetic at much smaller in-plane fields. In the case of VSTe, DMI is very small compared to the magnetic exchange $J$ and hence only SS appears without external field. The critical ordering temperature of VSTe is 780 K.

\section{Conclusion}
We have shown that reasonably strong DMIs can be obtained in Janus VXY monolayers by using first principle calculations, in spite of the relatively weak SOC of vanadium. This large value is directly related to the electric dipole induced by the dissimilar chalcogen elements, as already observed for the Rashba spin-splitting in Ref. \ref{Smaili2020}. In addition, we find that whereas VSSe possesses a weak out-of-plane anisotropy, VSeTe and VSTe are characterized by strong easy-plane anisotropy. This observation suggests that the two latter materials could be an interesting platform for spin superfluidity \cite{Takei2014}, although we leave this aspect to future work.

We emphasize that the DMI we obtain is rather weak (about one order of magnitude smaller than in transition metal multilayers \cite{Belabbes2016}), which is partially compensated by the fact that VXY is a monolayer, much thinner than traditional transition metal thin films. Therefore, homochiral spin spirals can be stabilized down to zero external field, and metastable skyrmions and bimerons can be obtained in VSeTe and VSSe, respectively. Nonetheless, the fairly large magnetic anisotropy of these monolayers prevents the stabilization of chiral textures without external magnetic field. We also emphasize that whereas our study focuses on stable ground states, the possibility to stabilize skyrmion and bimeron crystals paves the way to the realization of isolated metastable skyrmions and bimerons, which are of highest interest for potential applications.

This study, along with recent work focusing on different materials \cite{Liang2020,Yuan2020c}, suggests that transition metal dichalcogenides monolayers in Janus configuration can host a wealth of chiral textures in spite of their weak SOC. One can easily foresee that chemical surface engineering can also be favorably used to modulate the inversion symmetry breaking, which calls for further experimental exploration.

\acknowledgments
The work was supported by King Abdullah University of Science and Technology (KAUST) through the award OSR-2018-CRG7-3717 from the Office of Sponsored Research (OSR). This work used the resources of the Supercomputing Laboratory at King Abdullah University of Science and Technology (KAUST) in Thuwal, Saudi Arabia.

\bibliography{References}

\begin{thebibliography}{60}%
\makeatletter
\providecommand \@ifxundefined [1]{%
 \@ifx{#1\undefined}
}%
\providecommand \@ifnum [1]{%
 \ifnum #1\expandafter \@firstoftwo
 \else \expandafter \@secondoftwo
 \fi
}%
\providecommand \@ifx [1]{%
 \ifx #1\expandafter \@firstoftwo
 \else \expandafter \@secondoftwo
 \fi
}%
\providecommand \natexlab [1]{#1}%
\providecommand \enquote  [1]{``#1''}%
\providecommand \bibnamefont  [1]{#1}%
\providecommand \bibfnamefont [1]{#1}%
\providecommand \citenamefont [1]{#1}%
\providecommand \href@noop [0]{\@secondoftwo}%
\providecommand \href [0]{\begingroup \@sanitize@url \@href}%
\providecommand \@href[1]{\@@startlink{#1}\@@href}%
\providecommand \@@href[1]{\endgroup#1\@@endlink}%
\providecommand \@sanitize@url [0]{\catcode `\\12\catcode `\$12\catcode
  `\&12\catcode `\#12\catcode `\^12\catcode `\_12\catcode `\%12\relax}%
\providecommand \@@startlink[1]{}%
\providecommand \@@endlink[0]{}%
\providecommand \url  [0]{\begingroup\@sanitize@url \@url }%
\providecommand \@url [1]{\endgroup\@href {#1}{\urlprefix }}%
\providecommand \urlprefix  [0]{URL }%
\providecommand \Eprint [0]{\href }%
\providecommand \doibase [0]{http://dx.doi.org/}%
\providecommand \selectlanguage [0]{\@gobble}%
\providecommand \bibinfo  [0]{\@secondoftwo}%
\providecommand \bibfield  [0]{\@secondoftwo}%
\providecommand \translation [1]{[#1]}%
\providecommand \BibitemOpen [0]{}%
\providecommand \bibitemStop [0]{}%
\providecommand \bibitemNoStop [0]{.\EOS\space}%
\providecommand \EOS [0]{\spacefactor3000\relax}%
\providecommand \BibitemShut  [1]{\csname bibitem#1\endcsname}%
\let\auto@bib@innerbib\@empty
\bibitem [{\citenamefont {Nagaosa}\ and\ \citenamefont
  {Tokura}(2013)}]{nagaosa2013topological}%
  \BibitemOpen
  \bibfield  {author} {\bibinfo {author} {\bibfnamefont {N.}~\bibnamefont
  {Nagaosa}}\ and\ \bibinfo {author} {\bibfnamefont {Y.}~\bibnamefont
  {Tokura}},\ }\href {\doibase
  https://doi-org.libproxy.kaust.edu.sa/10.1038/nnano.2013.243} {\bibfield
  {journal} {\bibinfo  {journal} {Nature Nanotechnology}\ }\textbf {\bibinfo
  {volume} {8}},\ \bibinfo {pages} {899} (\bibinfo {year} {2013})}\BibitemShut
  {NoStop}%
\bibitem [{\citenamefont {Ryu}\ \emph {et~al.}(2013)\citenamefont {Ryu},
  \citenamefont {Thomas}, \citenamefont {Yang},\ and\ \citenamefont
  {Parkin}}]{ryu2013chiral}%
  \BibitemOpen
  \bibfield  {author} {\bibinfo {author} {\bibfnamefont {K.-S.}\ \bibnamefont
  {Ryu}}, \bibinfo {author} {\bibfnamefont {L.}~\bibnamefont {Thomas}},
  \bibinfo {author} {\bibfnamefont {S.-H.}\ \bibnamefont {Yang}}, \ and\
  \bibinfo {author} {\bibfnamefont {S.}~\bibnamefont {Parkin}},\ }\href
  {\doibase https://doi.org/10.1038/nnano.2013.102} {\bibfield  {journal}
  {\bibinfo  {journal} {Nature Nanotechnology}\ }\textbf {\bibinfo {volume}
  {8}},\ \bibinfo {pages} {527} (\bibinfo {year} {2013})}\BibitemShut {NoStop}%
\bibitem [{\citenamefont {Emori}\ \emph {et~al.}(2013)\citenamefont {Emori},
  \citenamefont {Bauer}, \citenamefont {Ahn}, \citenamefont {Martinez},\ and\
  \citenamefont {Beach}}]{emori2013current}%
  \BibitemOpen
  \bibfield  {author} {\bibinfo {author} {\bibfnamefont {S.}~\bibnamefont
  {Emori}}, \bibinfo {author} {\bibfnamefont {U.}~\bibnamefont {Bauer}},
  \bibinfo {author} {\bibfnamefont {S.-M.}\ \bibnamefont {Ahn}}, \bibinfo
  {author} {\bibfnamefont {E.}~\bibnamefont {Martinez}}, \ and\ \bibinfo
  {author} {\bibfnamefont {G.~S.}\ \bibnamefont {Beach}},\ }\href {\doibase
  https://doi.org/10.1038/nmat3675} {\bibfield  {journal} {\bibinfo  {journal}
  {Nature Materials}\ }\textbf {\bibinfo {volume} {12}},\ \bibinfo {pages}
  {611} (\bibinfo {year} {2013})}\BibitemShut {NoStop}%
\bibitem [{\citenamefont {Fert}\ \emph {et~al.}(2013)\citenamefont {Fert},
  \citenamefont {Cros},\ and\ \citenamefont {Sampaio}}]{fert2013skyrmions}%
  \BibitemOpen
  \bibfield  {author} {\bibinfo {author} {\bibfnamefont {A.}~\bibnamefont
  {Fert}}, \bibinfo {author} {\bibfnamefont {V.}~\bibnamefont {Cros}}, \ and\
  \bibinfo {author} {\bibfnamefont {J.}~\bibnamefont {Sampaio}},\ }\href
  {\doibase https://doi.org/10.1038/nnano.2013.29} {\bibfield  {journal}
  {\bibinfo  {journal} {Nature Nanotechnology}\ }\textbf {\bibinfo {volume}
  {8}},\ \bibinfo {pages} {152} (\bibinfo {year} {2013})}\BibitemShut {NoStop}%
\bibitem [{\citenamefont {Li}\ \emph {et~al.}(2017)\citenamefont {Li},
  \citenamefont {Kang}, \citenamefont {Huang}, \citenamefont {Zhang},\ and\
  \citenamefont {Zhou}}]{Li2017}%
  \BibitemOpen
  \bibfield  {author} {\bibinfo {author} {\bibfnamefont {S.}~\bibnamefont
  {Li}}, \bibinfo {author} {\bibfnamefont {W.}~\bibnamefont {Kang}}, \bibinfo
  {author} {\bibfnamefont {Y.}~\bibnamefont {Huang}}, \bibinfo {author}
  {\bibfnamefont {X.}~\bibnamefont {Zhang}}, \ and\ \bibinfo {author}
  {\bibfnamefont {Y.}~\bibnamefont {Zhou}},\ }\href@noop {} {\bibfield
  {journal} {\bibinfo  {journal} {Nanotechnology}\ }\textbf {\bibinfo {volume}
  {18}},\ \bibinfo {pages} {31LT01} (\bibinfo {year} {2017})}\BibitemShut
  {NoStop}%
\bibitem [{\citenamefont {Huang}\ \emph
  {et~al.}(2017{\natexlab{a}})\citenamefont {Huang}, \citenamefont {Kang},
  \citenamefont {Zhang},\ and\ \citenamefont {Zhou}}]{Huang2017c}%
  \BibitemOpen
  \bibfield  {author} {\bibinfo {author} {\bibfnamefont {Y.}~\bibnamefont
  {Huang}}, \bibinfo {author} {\bibfnamefont {W.}~\bibnamefont {Kang}},
  \bibinfo {author} {\bibfnamefont {X.}~\bibnamefont {Zhang}}, \ and\ \bibinfo
  {author} {\bibfnamefont {Y.}~\bibnamefont {Zhou}},\ }\href@noop {} {\bibfield
   {journal} {\bibinfo  {journal} {Nanotechnology}\ }\textbf {\bibinfo {volume}
  {28}},\ \bibinfo {pages} {08LT02} (\bibinfo {year}
  {2017}{\natexlab{a}})}\BibitemShut {NoStop}%
\bibitem [{\citenamefont {Prychynenko}\ \emph {et~al.}(2018)\citenamefont
  {Prychynenko}, \citenamefont {Sitte}, \citenamefont {Litzius}, \citenamefont
  {Kr{\"{u}}ger}, \citenamefont {Bourianoff}, \citenamefont {Kl{\"{a}}ui},
  \citenamefont {Sinova},\ and\ \citenamefont
  {Everschor-sitte}}]{Prychynenko2018}%
  \BibitemOpen
  \bibfield  {author} {\bibinfo {author} {\bibfnamefont {D.}~\bibnamefont
  {Prychynenko}}, \bibinfo {author} {\bibfnamefont {M.}~\bibnamefont {Sitte}},
  \bibinfo {author} {\bibfnamefont {K.}~\bibnamefont {Litzius}}, \bibinfo
  {author} {\bibfnamefont {B.}~\bibnamefont {Kr{\"{u}}ger}}, \bibinfo {author}
  {\bibfnamefont {G.}~\bibnamefont {Bourianoff}}, \bibinfo {author}
  {\bibfnamefont {M.}~\bibnamefont {Kl{\"{a}}ui}}, \bibinfo {author}
  {\bibfnamefont {J.}~\bibnamefont {Sinova}}, \ and\ \bibinfo {author}
  {\bibfnamefont {K.}~\bibnamefont {Everschor-sitte}},\ }\href {\doibase
  10.1103/PhysRevApplied.9.014034} {\bibfield  {journal} {\bibinfo  {journal}
  {Physical Review Applied}\ }\textbf {\bibinfo {volume} {9}},\ \bibinfo
  {pages} {14034} (\bibinfo {year} {2018})}\BibitemShut {NoStop}%
\bibitem [{\citenamefont {Luo}\ \emph {et~al.}(2018)\citenamefont {Luo},
  \citenamefont {Song}, \citenamefont {Li}, \citenamefont {Zhang},
  \citenamefont {Hong}, \citenamefont {Yang}, \citenamefont {Zou},
  \citenamefont {Xu},\ and\ \citenamefont {You}}]{Luo2018}%
  \BibitemOpen
  \bibfield  {author} {\bibinfo {author} {\bibfnamefont {S.}~\bibnamefont
  {Luo}}, \bibinfo {author} {\bibfnamefont {M.}~\bibnamefont {Song}}, \bibinfo
  {author} {\bibfnamefont {X.}~\bibnamefont {Li}}, \bibinfo {author}
  {\bibfnamefont {Y.}~\bibnamefont {Zhang}}, \bibinfo {author} {\bibfnamefont
  {J.}~\bibnamefont {Hong}}, \bibinfo {author} {\bibfnamefont {X.}~\bibnamefont
  {Yang}}, \bibinfo {author} {\bibfnamefont {X.}~\bibnamefont {Zou}}, \bibinfo
  {author} {\bibfnamefont {N.}~\bibnamefont {Xu}}, \ and\ \bibinfo {author}
  {\bibfnamefont {L.}~\bibnamefont {You}},\ }\href {\doibase
  10.1021/acs.nanolett.7b04722} {\bibfield  {journal} {\bibinfo  {journal}
  {Nano Letters}\ }\textbf {\bibinfo {volume} {18}},\ \bibinfo {pages} {1180}
  (\bibinfo {year} {2018})}\BibitemShut {NoStop}%
\bibitem [{\citenamefont {Z{\'{a}}zvorka}\ \emph {et~al.}(2019)\citenamefont
  {Z{\'{a}}zvorka}, \citenamefont {Jakobs}, \citenamefont {Heinze},
  \citenamefont {Keil}, \citenamefont {Kromin}, \citenamefont {Jaiswal},
  \citenamefont {Litzius}, \citenamefont {Jakob}, \citenamefont {Virnau},
  \citenamefont {Pinna}, \citenamefont {Everschor-sitte}, \citenamefont
  {R{\'{o}}zsa}, \citenamefont {Donges}, \citenamefont {Nowak},\ and\
  \citenamefont {Kl{\"{a}}ui}}]{Zazvorka2019}%
  \BibitemOpen
  \bibfield  {author} {\bibinfo {author} {\bibfnamefont {J.}~\bibnamefont
  {Z{\'{a}}zvorka}}, \bibinfo {author} {\bibfnamefont {F.}~\bibnamefont
  {Jakobs}}, \bibinfo {author} {\bibfnamefont {D.}~\bibnamefont {Heinze}},
  \bibinfo {author} {\bibfnamefont {N.}~\bibnamefont {Keil}}, \bibinfo {author}
  {\bibfnamefont {S.}~\bibnamefont {Kromin}}, \bibinfo {author} {\bibfnamefont
  {S.}~\bibnamefont {Jaiswal}}, \bibinfo {author} {\bibfnamefont
  {K.}~\bibnamefont {Litzius}}, \bibinfo {author} {\bibfnamefont
  {G.}~\bibnamefont {Jakob}}, \bibinfo {author} {\bibfnamefont
  {P.}~\bibnamefont {Virnau}}, \bibinfo {author} {\bibfnamefont
  {D.}~\bibnamefont {Pinna}}, \bibinfo {author} {\bibfnamefont
  {K.}~\bibnamefont {Everschor-sitte}}, \bibinfo {author} {\bibfnamefont
  {L.}~\bibnamefont {R{\'{o}}zsa}}, \bibinfo {author} {\bibfnamefont
  {A.}~\bibnamefont {Donges}}, \bibinfo {author} {\bibfnamefont
  {U.}~\bibnamefont {Nowak}}, \ and\ \bibinfo {author} {\bibfnamefont
  {M.}~\bibnamefont {Kl{\"{a}}ui}},\ }\href {\doibase
  10.1038/s41565-019-0436-8} {\bibfield  {journal} {\bibinfo  {journal} {Nature
  Nanotechnology}\ }\textbf {\bibinfo {volume} {14}},\ \bibinfo {pages} {658}
  (\bibinfo {year} {2019})}\BibitemShut {NoStop}%
\bibitem [{\citenamefont {Song}\ \emph {et~al.}(2020)\citenamefont {Song},
  \citenamefont {Jeong}, \citenamefont {Pan}, \citenamefont {Zhang},
  \citenamefont {Xia}, \citenamefont {Cha}, \citenamefont {Park}, \citenamefont
  {Kim}, \citenamefont {Finizio}, \citenamefont {Raabe}, \citenamefont {Chang},
  \citenamefont {Zhou}, \citenamefont {Zhao}, \citenamefont {Kang},
  \citenamefont {Ju},\ and\ \citenamefont {Woo}}]{Song2020}%
  \BibitemOpen
  \bibfield  {author} {\bibinfo {author} {\bibfnamefont {K.~M.}\ \bibnamefont
  {Song}}, \bibinfo {author} {\bibfnamefont {J.-s.}\ \bibnamefont {Jeong}},
  \bibinfo {author} {\bibfnamefont {B.}~\bibnamefont {Pan}}, \bibinfo {author}
  {\bibfnamefont {X.}~\bibnamefont {Zhang}}, \bibinfo {author} {\bibfnamefont
  {J.}~\bibnamefont {Xia}}, \bibinfo {author} {\bibfnamefont {S.}~\bibnamefont
  {Cha}}, \bibinfo {author} {\bibfnamefont {T.-e.}\ \bibnamefont {Park}},
  \bibinfo {author} {\bibfnamefont {K.}~\bibnamefont {Kim}}, \bibinfo {author}
  {\bibfnamefont {S.}~\bibnamefont {Finizio}}, \bibinfo {author} {\bibfnamefont
  {J.}~\bibnamefont {Raabe}}, \bibinfo {author} {\bibfnamefont
  {J.}~\bibnamefont {Chang}}, \bibinfo {author} {\bibfnamefont
  {Y.}~\bibnamefont {Zhou}}, \bibinfo {author} {\bibfnamefont {W.}~\bibnamefont
  {Zhao}}, \bibinfo {author} {\bibfnamefont {W.}~\bibnamefont {Kang}}, \bibinfo
  {author} {\bibfnamefont {H.}~\bibnamefont {Ju}}, \ and\ \bibinfo {author}
  {\bibfnamefont {S.}~\bibnamefont {Woo}},\ }\href {\doibase
  10.1038/s41928-020-0385-0} {\bibfield  {journal} {\bibinfo  {journal} {Nature
  Electronics}\ }\textbf {\bibinfo {volume} {3}},\ \bibinfo {pages} {148}
  (\bibinfo {year} {2020})}\BibitemShut {NoStop}%
\bibitem [{\citenamefont {Moriya}(1960)}]{Moriya1960anisotropic}%
  \BibitemOpen
  \bibfield  {author} {\bibinfo {author} {\bibfnamefont {T.}~\bibnamefont
  {Moriya}},\ }\href {\doibase https://doi.org/10.1103/PhysRev.120.91}
  {\bibfield  {journal} {\bibinfo  {journal} {Physical Review}\ }\textbf
  {\bibinfo {volume} {120}},\ \bibinfo {pages} {91} (\bibinfo {year}
  {1960})}\BibitemShut {NoStop}%
\bibitem [{\citenamefont {Dzyaloshinskii}(1957)}]{Dzyaloshinskii1957}%
  \BibitemOpen
  \bibfield  {author} {\bibinfo {author} {\bibfnamefont {I.}~\bibnamefont
  {Dzyaloshinskii}},\ }\href@noop {} {\bibfield  {journal} {\bibinfo  {journal}
  {Soviet Physics JETP}\ }\textbf {\bibinfo {volume} {5}},\ \bibinfo {pages}
  {1259} (\bibinfo {year} {1957})}\BibitemShut {NoStop}%
\bibitem [{\citenamefont {M{\"{u}}hlbauer}\ \emph {et~al.}(2009)\citenamefont
  {M{\"{u}}hlbauer}, \citenamefont {Binz}, \citenamefont {Jonietz},
  \citenamefont {Pfleiderer}, \citenamefont {Rosch}, \citenamefont {Neubauer},
  \citenamefont {Georgii},\ and\ \citenamefont {B{\"{o}}ni}}]{Muhlbauer2009}%
  \BibitemOpen
  \bibfield  {author} {\bibinfo {author} {\bibfnamefont {S.}~\bibnamefont
  {M{\"{u}}hlbauer}}, \bibinfo {author} {\bibfnamefont {B.}~\bibnamefont
  {Binz}}, \bibinfo {author} {\bibfnamefont {F.}~\bibnamefont {Jonietz}},
  \bibinfo {author} {\bibfnamefont {C.}~\bibnamefont {Pfleiderer}}, \bibinfo
  {author} {\bibfnamefont {A.}~\bibnamefont {Rosch}}, \bibinfo {author}
  {\bibfnamefont {A.}~\bibnamefont {Neubauer}}, \bibinfo {author}
  {\bibfnamefont {R.}~\bibnamefont {Georgii}}, \ and\ \bibinfo {author}
  {\bibfnamefont {P.}~\bibnamefont {B{\"{o}}ni}},\ }\href
  {http://www.sciencemag.org/content/323/5916/915.short} {\bibfield  {journal}
  {\bibinfo  {journal} {Science (New York, N.Y.)}\ }\textbf {\bibinfo {volume}
  {323}},\ \bibinfo {pages} {915} (\bibinfo {year} {2009})}\BibitemShut
  {NoStop}%
\bibitem [{\citenamefont {Yu}\ \emph {et~al.}(2010)\citenamefont {Yu},
  \citenamefont {Onose}, \citenamefont {Kanazawa}, \citenamefont {Park},
  \citenamefont {Han}, \citenamefont {Matsui}, \citenamefont {Nagaosa},\ and\
  \citenamefont {Tokura}}]{Yu2010}%
  \BibitemOpen
  \bibfield  {author} {\bibinfo {author} {\bibfnamefont {X.~Z.}\ \bibnamefont
  {Yu}}, \bibinfo {author} {\bibfnamefont {Y.}~\bibnamefont {Onose}}, \bibinfo
  {author} {\bibfnamefont {N.}~\bibnamefont {Kanazawa}}, \bibinfo {author}
  {\bibfnamefont {J.~H.}\ \bibnamefont {Park}}, \bibinfo {author}
  {\bibfnamefont {J.~H.}\ \bibnamefont {Han}}, \bibinfo {author} {\bibfnamefont
  {Y.}~\bibnamefont {Matsui}}, \bibinfo {author} {\bibfnamefont
  {N.}~\bibnamefont {Nagaosa}}, \ and\ \bibinfo {author} {\bibfnamefont
  {Y.}~\bibnamefont {Tokura}},\ }\href {\doibase 10.1038/nature09124}
  {\bibfield  {journal} {\bibinfo  {journal} {Nature}\ }\textbf {\bibinfo
  {volume} {465}},\ \bibinfo {pages} {901} (\bibinfo {year}
  {2010})}\BibitemShut {NoStop}%
\bibitem [{\citenamefont {Jiang}\ \emph {et~al.}(2015)\citenamefont {Jiang},
  \citenamefont {Upadhyaya}, \citenamefont {Zhang}, \citenamefont {Yu},
  \citenamefont {Jungfleisch}, \citenamefont {Fradin}, \citenamefont {Pearson},
  \citenamefont {Tserkovnyak}, \citenamefont {Wang}, \citenamefont {Heinonen},
  \citenamefont {Velthuis},\ and\ \citenamefont {Hoffmann}}]{Jiang2015}%
  \BibitemOpen
  \bibfield  {author} {\bibinfo {author} {\bibfnamefont {W.}~\bibnamefont
  {Jiang}}, \bibinfo {author} {\bibfnamefont {P.}~\bibnamefont {Upadhyaya}},
  \bibinfo {author} {\bibfnamefont {W.}~\bibnamefont {Zhang}}, \bibinfo
  {author} {\bibfnamefont {G.}~\bibnamefont {Yu}}, \bibinfo {author}
  {\bibfnamefont {M.~B.}\ \bibnamefont {Jungfleisch}}, \bibinfo {author}
  {\bibfnamefont {F.~Y.}\ \bibnamefont {Fradin}}, \bibinfo {author}
  {\bibfnamefont {J.~E.}\ \bibnamefont {Pearson}}, \bibinfo {author}
  {\bibfnamefont {Y.}~\bibnamefont {Tserkovnyak}}, \bibinfo {author}
  {\bibfnamefont {K.~L.}\ \bibnamefont {Wang}}, \bibinfo {author}
  {\bibfnamefont {O.}~\bibnamefont {Heinonen}}, \bibinfo {author}
  {\bibfnamefont {S.~G.~E.}\ \bibnamefont {Velthuis}}, \ and\ \bibinfo {author}
  {\bibfnamefont {A.}~\bibnamefont {Hoffmann}},\ }\href@noop {} {\bibfield
  {journal} {\bibinfo  {journal} {Science}\ }\textbf {\bibinfo {volume}
  {349}},\ \bibinfo {pages} {283} (\bibinfo {year} {2015})}\BibitemShut
  {NoStop}%
\bibitem [{\citenamefont {Chen}\ \emph {et~al.}(2015)\citenamefont {Chen},
  \citenamefont {Mascaraque}, \citenamefont {N'Diaye},\ and\ \citenamefont
  {Schmid}}]{Chen2015b}%
  \BibitemOpen
  \bibfield  {author} {\bibinfo {author} {\bibfnamefont {G.}~\bibnamefont
  {Chen}}, \bibinfo {author} {\bibfnamefont {A.}~\bibnamefont {Mascaraque}},
  \bibinfo {author} {\bibfnamefont {A.~T.}\ \bibnamefont {N'Diaye}}, \ and\
  \bibinfo {author} {\bibfnamefont {A.~K.}\ \bibnamefont {Schmid}},\ }\href
  {\doibase 10.1063/1.4922726} {\bibfield  {journal} {\bibinfo  {journal}
  {Applied Physics Letters}\ }\textbf {\bibinfo {volume} {106}},\ \bibinfo
  {pages} {242404} (\bibinfo {year} {2015})}\BibitemShut {NoStop}%
\bibitem [{\citenamefont {Boulle}\ \emph {et~al.}(2016)\citenamefont {Boulle},
  \citenamefont {Vogel}, \citenamefont {Yang}, \citenamefont {Pizzini},
  \citenamefont {Chaves}, \citenamefont {Locatelli}, \citenamefont {Sala},
  \citenamefont {Buda-Prejbeanu}, \citenamefont {Klein}, \citenamefont
  {Belmeguenai}, \citenamefont {Roussign{\'{e}}}, \citenamefont {Stashkevich},
  \citenamefont {Ch{\'{e}}rif}, \citenamefont {Aballe}, \citenamefont
  {Foerster}, \citenamefont {Chshiev}, \citenamefont {Auffret}, \citenamefont
  {Miron},\ and\ \citenamefont {Gaudin}}]{Boulle2016}%
  \BibitemOpen
  \bibfield  {author} {\bibinfo {author} {\bibfnamefont {O.}~\bibnamefont
  {Boulle}}, \bibinfo {author} {\bibfnamefont {J.}~\bibnamefont {Vogel}},
  \bibinfo {author} {\bibfnamefont {H.}~\bibnamefont {Yang}}, \bibinfo {author}
  {\bibfnamefont {S.}~\bibnamefont {Pizzini}}, \bibinfo {author} {\bibfnamefont
  {D.~d.~S.}\ \bibnamefont {Chaves}}, \bibinfo {author} {\bibfnamefont
  {A.}~\bibnamefont {Locatelli}}, \bibinfo {author} {\bibfnamefont {T.~O.
  M.~A.}\ \bibnamefont {Sala}}, \bibinfo {author} {\bibfnamefont {L.~D.}\
  \bibnamefont {Buda-Prejbeanu}}, \bibinfo {author} {\bibfnamefont
  {O.}~\bibnamefont {Klein}}, \bibinfo {author} {\bibfnamefont
  {M.}~\bibnamefont {Belmeguenai}}, \bibinfo {author} {\bibfnamefont
  {Y.}~\bibnamefont {Roussign{\'{e}}}}, \bibinfo {author} {\bibfnamefont
  {A.}~\bibnamefont {Stashkevich}}, \bibinfo {author} {\bibfnamefont {S.~M.}\
  \bibnamefont {Ch{\'{e}}rif}}, \bibinfo {author} {\bibfnamefont
  {L.}~\bibnamefont {Aballe}}, \bibinfo {author} {\bibfnamefont
  {M.}~\bibnamefont {Foerster}}, \bibinfo {author} {\bibfnamefont
  {M.}~\bibnamefont {Chshiev}}, \bibinfo {author} {\bibfnamefont
  {S.}~\bibnamefont {Auffret}}, \bibinfo {author} {\bibfnamefont {I.~M.}\
  \bibnamefont {Miron}}, \ and\ \bibinfo {author} {\bibfnamefont
  {G.}~\bibnamefont {Gaudin}},\ }\href {\doibase 10.1038/NNANO.2015.315}
  {\bibfield  {journal} {\bibinfo  {journal} {Nature Nanotechnology}\ }\textbf
  {\bibinfo {volume} {11}},\ \bibinfo {pages} {449} (\bibinfo {year}
  {2016})}\BibitemShut {NoStop}%
\bibitem [{\citenamefont {Moreau-Luchaire}\ \emph {et~al.}(2016)\citenamefont
  {Moreau-Luchaire}, \citenamefont {Moutaﬁs}, \citenamefont {Reyren},
  \citenamefont {Sampaio}, \citenamefont {Vaz}, \citenamefont {{Van Horne}},
  \citenamefont {Bouzehouane}, \citenamefont {Garcia}, \citenamefont
  {Deranlot}, \citenamefont {Warnicke}, \citenamefont {Wohlh{\"{u}}ter},
  \citenamefont {George}, \citenamefont {Weigand}, \citenamefont {Raabe},
  \citenamefont {Cros},\ and\ \citenamefont {Fert}}]{Moreau-Luchaire2016}%
  \BibitemOpen
  \bibfield  {author} {\bibinfo {author} {\bibfnamefont {C.}~\bibnamefont
  {Moreau-Luchaire}}, \bibinfo {author} {\bibfnamefont {C.}~\bibnamefont
  {Moutaﬁs}}, \bibinfo {author} {\bibfnamefont {N.}~\bibnamefont {Reyren}},
  \bibinfo {author} {\bibfnamefont {J.}~\bibnamefont {Sampaio}}, \bibinfo
  {author} {\bibfnamefont {C.~A.~F.}\ \bibnamefont {Vaz}}, \bibinfo {author}
  {\bibfnamefont {N.}~\bibnamefont {{Van Horne}}}, \bibinfo {author}
  {\bibfnamefont {K.}~\bibnamefont {Bouzehouane}}, \bibinfo {author}
  {\bibfnamefont {K.}~\bibnamefont {Garcia}}, \bibinfo {author} {\bibfnamefont
  {C.}~\bibnamefont {Deranlot}}, \bibinfo {author} {\bibfnamefont
  {P.}~\bibnamefont {Warnicke}}, \bibinfo {author} {\bibfnamefont
  {P.}~\bibnamefont {Wohlh{\"{u}}ter}}, \bibinfo {author} {\bibfnamefont
  {J.-M.}\ \bibnamefont {George}}, \bibinfo {author} {\bibfnamefont
  {M.}~\bibnamefont {Weigand}}, \bibinfo {author} {\bibfnamefont
  {J.}~\bibnamefont {Raabe}}, \bibinfo {author} {\bibfnamefont
  {V.}~\bibnamefont {Cros}}, \ and\ \bibinfo {author} {\bibfnamefont
  {A.}~\bibnamefont {Fert}},\ }\href {\doibase 10.1038/nnano.2015.313}
  {\bibfield  {journal} {\bibinfo  {journal} {Nature Nanotechnology}\ }\textbf
  {\bibinfo {volume} {11}},\ \bibinfo {pages} {444} (\bibinfo {year}
  {2016})}\BibitemShut {NoStop}%
\bibitem [{\citenamefont {Woo}\ \emph {et~al.}(2016)\citenamefont {Woo},
  \citenamefont {Litzius}, \citenamefont {Kr{\"{u}}ger}, \citenamefont {Im},
  \citenamefont {Caretta}, \citenamefont {Richter}, \citenamefont {Mann},
  \citenamefont {Krone}, \citenamefont {Reeve}, \citenamefont {Weigand},
  \citenamefont {Agrawal}, \citenamefont {Lemesh}, \citenamefont {Mawass},
  \citenamefont {Fischer}, \citenamefont {Kl{\"{a}}ui},\ and\ \citenamefont
  {Beach}}]{Woo2016}%
  \BibitemOpen
  \bibfield  {author} {\bibinfo {author} {\bibfnamefont {S.}~\bibnamefont
  {Woo}}, \bibinfo {author} {\bibfnamefont {K.}~\bibnamefont {Litzius}},
  \bibinfo {author} {\bibfnamefont {B.}~\bibnamefont {Kr{\"{u}}ger}}, \bibinfo
  {author} {\bibfnamefont {M.-Y.}\ \bibnamefont {Im}}, \bibinfo {author}
  {\bibfnamefont {L.}~\bibnamefont {Caretta}}, \bibinfo {author} {\bibfnamefont
  {K.}~\bibnamefont {Richter}}, \bibinfo {author} {\bibfnamefont
  {M.}~\bibnamefont {Mann}}, \bibinfo {author} {\bibfnamefont {A.}~\bibnamefont
  {Krone}}, \bibinfo {author} {\bibfnamefont {R.~M.}\ \bibnamefont {Reeve}},
  \bibinfo {author} {\bibfnamefont {M.}~\bibnamefont {Weigand}}, \bibinfo
  {author} {\bibfnamefont {P.}~\bibnamefont {Agrawal}}, \bibinfo {author}
  {\bibfnamefont {I.}~\bibnamefont {Lemesh}}, \bibinfo {author} {\bibfnamefont
  {M.-A.}\ \bibnamefont {Mawass}}, \bibinfo {author} {\bibfnamefont
  {P.}~\bibnamefont {Fischer}}, \bibinfo {author} {\bibfnamefont
  {M.}~\bibnamefont {Kl{\"{a}}ui}}, \ and\ \bibinfo {author} {\bibfnamefont
  {G.~S.~D.}\ \bibnamefont {Beach}},\ }\href {\doibase 10.1038/nmat4593}
  {\bibfield  {journal} {\bibinfo  {journal} {Nature Materials}\ }\textbf
  {\bibinfo {volume} {15}},\ \bibinfo {pages} {501} (\bibinfo {year}
  {2016})}\BibitemShut {NoStop}%
\bibitem [{\citenamefont {Gong}\ \emph {et~al.}(2017)\citenamefont {Gong},
  \citenamefont {Li}, \citenamefont {Li}, \citenamefont {Ji}, \citenamefont
  {Stern}, \citenamefont {Xia}, \citenamefont {Cao}, \citenamefont {Bao},
  \citenamefont {Wang}, \citenamefont {Wang}, \citenamefont {Qiu},
  \citenamefont {Cava}, \citenamefont {Louie}, \citenamefont {Xia},\ and\
  \citenamefont {Zhang}}]{Gong2017}%
  \BibitemOpen
  \bibfield  {author} {\bibinfo {author} {\bibfnamefont {C.}~\bibnamefont
  {Gong}}, \bibinfo {author} {\bibfnamefont {L.}~\bibnamefont {Li}}, \bibinfo
  {author} {\bibfnamefont {Z.}~\bibnamefont {Li}}, \bibinfo {author}
  {\bibfnamefont {H.}~\bibnamefont {Ji}}, \bibinfo {author} {\bibfnamefont
  {A.}~\bibnamefont {Stern}}, \bibinfo {author} {\bibfnamefont
  {Y.}~\bibnamefont {Xia}}, \bibinfo {author} {\bibfnamefont {T.}~\bibnamefont
  {Cao}}, \bibinfo {author} {\bibfnamefont {W.}~\bibnamefont {Bao}}, \bibinfo
  {author} {\bibfnamefont {C.}~\bibnamefont {Wang}}, \bibinfo {author}
  {\bibfnamefont {Y.}~\bibnamefont {Wang}}, \bibinfo {author} {\bibfnamefont
  {Z.}~\bibnamefont {Qiu}}, \bibinfo {author} {\bibfnamefont {R.}~\bibnamefont
  {Cava}}, \bibinfo {author} {\bibfnamefont {S.}~\bibnamefont {Louie}},
  \bibinfo {author} {\bibfnamefont {J.}~\bibnamefont {Xia}}, \ and\ \bibinfo
  {author} {\bibfnamefont {X.}~\bibnamefont {Zhang}},\ }\href {\doibase
  https://doi.org/10.1038/nature22060} {\bibfield  {journal} {\bibinfo
  {journal} {Nature}\ }\textbf {\bibinfo {volume} {546}},\ \bibinfo {pages}
  {265} (\bibinfo {year} {2017})}\BibitemShut {NoStop}%
\bibitem [{\citenamefont {O’Hara}\ \emph {et~al.}(2018)\citenamefont
  {O’Hara}, \citenamefont {Zhu}, \citenamefont {Trout}, \citenamefont
  {Ahmed}, \citenamefont {Luo}, \citenamefont {Lee}, \citenamefont {Brenner},
  \citenamefont {Rajan}, \citenamefont {Gupta}, \citenamefont {McComb} \emph
  {et~al.}}]{o2018room}%
  \BibitemOpen
  \bibfield  {author} {\bibinfo {author} {\bibfnamefont {D.~J.}\ \bibnamefont
  {O’Hara}}, \bibinfo {author} {\bibfnamefont {T.}~\bibnamefont {Zhu}},
  \bibinfo {author} {\bibfnamefont {A.~H.}\ \bibnamefont {Trout}}, \bibinfo
  {author} {\bibfnamefont {A.~S.}\ \bibnamefont {Ahmed}}, \bibinfo {author}
  {\bibfnamefont {Y.~K.}\ \bibnamefont {Luo}}, \bibinfo {author} {\bibfnamefont
  {C.~H.}\ \bibnamefont {Lee}}, \bibinfo {author} {\bibfnamefont {M.~R.}\
  \bibnamefont {Brenner}}, \bibinfo {author} {\bibfnamefont {S.}~\bibnamefont
  {Rajan}}, \bibinfo {author} {\bibfnamefont {J.~A.}\ \bibnamefont {Gupta}},
  \bibinfo {author} {\bibfnamefont {D.~W.}\ \bibnamefont {McComb}},  \emph
  {et~al.},\ }\href {\doibase https://doi.org/10.1021/acs.nanolett.8b00683}
  {\bibfield  {journal} {\bibinfo  {journal} {Nano Letters}\ }\textbf {\bibinfo
  {volume} {18}},\ \bibinfo {pages} {3125} (\bibinfo {year}
  {2018})}\BibitemShut {NoStop}%
\bibitem [{\citenamefont {O'Hara}\ \emph
  {et~al.}(2018{\natexlab{a}})\citenamefont {O'Hara}, \citenamefont {Zhu},
  \citenamefont {Trout}, \citenamefont {Ahmed}, \citenamefont {Luo},
  \citenamefont {Lee}, \citenamefont {Brenner}, \citenamefont {Rajan},
  \citenamefont {Gupta}, \citenamefont {McComb},\ and\ \citenamefont
  {Kawakami}}]{O'Hara2018}%
  \BibitemOpen
  \bibfield  {author} {\bibinfo {author} {\bibfnamefont {D.~J.}\ \bibnamefont
  {O'Hara}}, \bibinfo {author} {\bibfnamefont {T.}~\bibnamefont {Zhu}},
  \bibinfo {author} {\bibfnamefont {A.~H.}\ \bibnamefont {Trout}}, \bibinfo
  {author} {\bibfnamefont {A.~S.}\ \bibnamefont {Ahmed}}, \bibinfo {author}
  {\bibfnamefont {Y.~K.}\ \bibnamefont {Luo}}, \bibinfo {author} {\bibfnamefont
  {C.~H.}\ \bibnamefont {Lee}}, \bibinfo {author} {\bibfnamefont {M.~R.}\
  \bibnamefont {Brenner}}, \bibinfo {author} {\bibfnamefont {S.}~\bibnamefont
  {Rajan}}, \bibinfo {author} {\bibfnamefont {J.~A.}\ \bibnamefont {Gupta}},
  \bibinfo {author} {\bibfnamefont {D.~W.}\ \bibnamefont {McComb}}, \ and\
  \bibinfo {author} {\bibfnamefont {R.~K.}\ \bibnamefont {Kawakami}},\ }\href
  {\doibase https://doi.org/10.1021/acs.nanolett.8b00683} {\bibfield  {journal}
  {\bibinfo  {journal} {Nano Letters}\ }\textbf {\bibinfo {volume} {18}},\
  \bibinfo {pages} {3125} (\bibinfo {year} {2018}{\natexlab{a}})}\BibitemShut
  {NoStop}%
\bibitem [{\citenamefont {O'Hara}\ \emph
  {et~al.}(2018{\natexlab{b}})\citenamefont {O'Hara}, \citenamefont {Zhu},\
  and\ \citenamefont {Kawakami}}]{O'Hara2018b}%
  \BibitemOpen
  \bibfield  {author} {\bibinfo {author} {\bibfnamefont {D.~J.}\ \bibnamefont
  {O'Hara}}, \bibinfo {author} {\bibfnamefont {T.}~\bibnamefont {Zhu}}, \ and\
  \bibinfo {author} {\bibfnamefont {R.~K.}\ \bibnamefont {Kawakami}},\ }\href
  {\doibase https://doi.org/10.1109/LMAG.2018.2867339} {\bibfield  {journal}
  {\bibinfo  {journal} {IEEE Magnetics Letters}\ }\textbf {\bibinfo {volume}
  {9}},\ \bibinfo {pages} {1} (\bibinfo {year}
  {2018}{\natexlab{b}})}\BibitemShut {NoStop}%
\bibitem [{\citenamefont {Bonilla}\ \emph {et~al.}(2018)\citenamefont
  {Bonilla}, \citenamefont {Kolekar}, \citenamefont {Ma}, \citenamefont {Diaz},
  \citenamefont {Kalappattil}, \citenamefont {Das}, \citenamefont {Eggers},
  \citenamefont {Gutierrez}, \citenamefont {Phan},\ and\ \citenamefont
  {Batzill}}]{bonilla2018strong}%
  \BibitemOpen
  \bibfield  {author} {\bibinfo {author} {\bibfnamefont {M.}~\bibnamefont
  {Bonilla}}, \bibinfo {author} {\bibfnamefont {S.}~\bibnamefont {Kolekar}},
  \bibinfo {author} {\bibfnamefont {Y.}~\bibnamefont {Ma}}, \bibinfo {author}
  {\bibfnamefont {H.~C.}\ \bibnamefont {Diaz}}, \bibinfo {author}
  {\bibfnamefont {V.}~\bibnamefont {Kalappattil}}, \bibinfo {author}
  {\bibfnamefont {R.}~\bibnamefont {Das}}, \bibinfo {author} {\bibfnamefont
  {T.}~\bibnamefont {Eggers}}, \bibinfo {author} {\bibfnamefont {H.~R.}\
  \bibnamefont {Gutierrez}}, \bibinfo {author} {\bibfnamefont {M.-H.}\
  \bibnamefont {Phan}}, \ and\ \bibinfo {author} {\bibfnamefont
  {M.}~\bibnamefont {Batzill}},\ }\href
  {https://www-nature-com.libproxy.kaust.edu.sa/articles/s41565-018-0063-9.pdf}
  {\bibfield  {journal} {\bibinfo  {journal} {Nature Nanotechnology}\ }\textbf
  {\bibinfo {volume} {13}},\ \bibinfo {pages} {289} (\bibinfo {year}
  {2018})}\BibitemShut {NoStop}%
\bibitem [{\citenamefont {Huang}\ \emph
  {et~al.}(2017{\natexlab{b}})\citenamefont {Huang}, \citenamefont {Clark},
  \citenamefont {Navarro-Moratalla}, \citenamefont {Klein}, \citenamefont
  {Cheng}, \citenamefont {Seyler}, \citenamefont {Zhong}, \citenamefont
  {Schmidgall}, \citenamefont {McGuire}, \citenamefont {Cobden} \emph
  {et~al.}}]{huang2017layer}%
  \BibitemOpen
  \bibfield  {author} {\bibinfo {author} {\bibfnamefont {B.}~\bibnamefont
  {Huang}}, \bibinfo {author} {\bibfnamefont {G.}~\bibnamefont {Clark}},
  \bibinfo {author} {\bibfnamefont {E.}~\bibnamefont {Navarro-Moratalla}},
  \bibinfo {author} {\bibfnamefont {D.~R.}\ \bibnamefont {Klein}}, \bibinfo
  {author} {\bibfnamefont {R.}~\bibnamefont {Cheng}}, \bibinfo {author}
  {\bibfnamefont {K.~L.}\ \bibnamefont {Seyler}}, \bibinfo {author}
  {\bibfnamefont {D.}~\bibnamefont {Zhong}}, \bibinfo {author} {\bibfnamefont
  {E.}~\bibnamefont {Schmidgall}}, \bibinfo {author} {\bibfnamefont {M.~A.}\
  \bibnamefont {McGuire}}, \bibinfo {author} {\bibfnamefont {D.~H.}\
  \bibnamefont {Cobden}},  \emph {et~al.},\ }\href {\doibase
  https://arxiv.org/ct?url=https%3A%2F%2Fdx.doi.org%2F10.1038%2Fnature22391&v=8dcccde4}
  {\bibfield  {journal} {\bibinfo  {journal} {Nature}\ }\textbf {\bibinfo
  {volume} {546}},\ \bibinfo {pages} {270} (\bibinfo {year}
  {2017}{\natexlab{b}})}\BibitemShut {NoStop}%
\bibitem [{\citenamefont {Fei}\ \emph {et~al.}(2018)\citenamefont {Fei},
  \citenamefont {Huang}, \citenamefont {Malinowski}, \citenamefont {Wang},
  \citenamefont {Song}, \citenamefont {Sanchez}, \citenamefont {Yao},
  \citenamefont {Xiao}, \citenamefont {Zhu}, \citenamefont {May} \emph
  {et~al.}}]{fei2018two}%
  \BibitemOpen
  \bibfield  {author} {\bibinfo {author} {\bibfnamefont {Z.}~\bibnamefont
  {Fei}}, \bibinfo {author} {\bibfnamefont {B.}~\bibnamefont {Huang}}, \bibinfo
  {author} {\bibfnamefont {P.}~\bibnamefont {Malinowski}}, \bibinfo {author}
  {\bibfnamefont {W.}~\bibnamefont {Wang}}, \bibinfo {author} {\bibfnamefont
  {T.}~\bibnamefont {Song}}, \bibinfo {author} {\bibfnamefont {J.}~\bibnamefont
  {Sanchez}}, \bibinfo {author} {\bibfnamefont {W.}~\bibnamefont {Yao}},
  \bibinfo {author} {\bibfnamefont {D.}~\bibnamefont {Xiao}}, \bibinfo {author}
  {\bibfnamefont {X.}~\bibnamefont {Zhu}}, \bibinfo {author} {\bibfnamefont
  {A.~F.}\ \bibnamefont {May}},  \emph {et~al.},\ }\href
  {https://www-nature-com.libproxy.kaust.edu.sa/articles/s41563-018-0149-7.pdf}
  {\bibfield  {journal} {\bibinfo  {journal} {Nature Materials}\ }\textbf
  {\bibinfo {volume} {17}},\ \bibinfo {pages} {778} (\bibinfo {year}
  {2018})}\BibitemShut {NoStop}%
\bibitem [{\citenamefont {Cheng}\ \emph {et~al.}(2013)\citenamefont {Cheng},
  \citenamefont {Zhu}, \citenamefont {Tahir},\ and\ \citenamefont
  {Schwingenschl{\"{o}}gl}}]{Cheng2013b}%
  \BibitemOpen
  \bibfield  {author} {\bibinfo {author} {\bibfnamefont {Y.~C.}\ \bibnamefont
  {Cheng}}, \bibinfo {author} {\bibfnamefont {Z.~Y.}\ \bibnamefont {Zhu}},
  \bibinfo {author} {\bibfnamefont {M.}~\bibnamefont {Tahir}}, \ and\ \bibinfo
  {author} {\bibfnamefont {U.}~\bibnamefont {Schwingenschl{\"{o}}gl}},\ }\href
  {\doibase 10.1209/0295-5075/102/57001} {\bibfield  {journal} {\bibinfo
  {journal} {Europhysics Letters}\ }\textbf {\bibinfo {volume} {102}},\
  \bibinfo {pages} {57001} (\bibinfo {year} {2013})}\BibitemShut {NoStop}%
\bibitem [{\citenamefont {Lu}\ \emph {et~al.}(2017)\citenamefont {Lu},
  \citenamefont {Zhu}, \citenamefont {Xiao}, \citenamefont {Chuu},
  \citenamefont {Han}, \citenamefont {Chiu}, \citenamefont {Cheng},
  \citenamefont {Yang}, \citenamefont {Wei}, \citenamefont {Yang} \emph
  {et~al.}}]{lu2017janus}%
  \BibitemOpen
  \bibfield  {author} {\bibinfo {author} {\bibfnamefont {A.-Y.}\ \bibnamefont
  {Lu}}, \bibinfo {author} {\bibfnamefont {H.}~\bibnamefont {Zhu}}, \bibinfo
  {author} {\bibfnamefont {J.}~\bibnamefont {Xiao}}, \bibinfo {author}
  {\bibfnamefont {C.-P.}\ \bibnamefont {Chuu}}, \bibinfo {author}
  {\bibfnamefont {Y.}~\bibnamefont {Han}}, \bibinfo {author} {\bibfnamefont
  {M.-H.}\ \bibnamefont {Chiu}}, \bibinfo {author} {\bibfnamefont {C.-C.}\
  \bibnamefont {Cheng}}, \bibinfo {author} {\bibfnamefont {C.-W.}\ \bibnamefont
  {Yang}}, \bibinfo {author} {\bibfnamefont {K.-H.}\ \bibnamefont {Wei}},
  \bibinfo {author} {\bibfnamefont {Y.}~\bibnamefont {Yang}},  \emph {et~al.},\
  }\href
  {https://www-nature-com.libproxy.kaust.edu.sa/articles/nnano.2017.100.pdf}
  {\bibfield  {journal} {\bibinfo  {journal} {Nature Nanotechnology}\ }\textbf
  {\bibinfo {volume} {12}},\ \bibinfo {pages} {744} (\bibinfo {year}
  {2017})}\BibitemShut {NoStop}%
\bibitem [{\citenamefont {Zhang}\ \emph {et~al.}(2017)\citenamefont {Zhang},
  \citenamefont {Jia}, \citenamefont {Kholmanov}, \citenamefont {Dong},
  \citenamefont {Er}, \citenamefont {Chen},\ and\ \citenamefont
  {Guo}}]{Zhang2017}%
  \BibitemOpen
  \bibfield  {author} {\bibinfo {author} {\bibfnamefont {J.}~\bibnamefont
  {Zhang}}, \bibinfo {author} {\bibfnamefont {S.}~\bibnamefont {Jia}}, \bibinfo
  {author} {\bibfnamefont {I.}~\bibnamefont {Kholmanov}}, \bibinfo {author}
  {\bibfnamefont {L.}~\bibnamefont {Dong}}, \bibinfo {author} {\bibfnamefont
  {D.}~\bibnamefont {Er}}, \bibinfo {author} {\bibfnamefont {W.}~\bibnamefont
  {Chen}}, \ and\ \bibinfo {author} {\bibfnamefont {H.}~\bibnamefont {Guo}},\
  }\href {\doibase 10.1021/acsnano.7b03186} {\bibfield  {journal} {\bibinfo
  {journal} {ACS Nano}\ }\textbf {\bibinfo {volume} {11}},\ \bibinfo {pages}
  {8192} (\bibinfo {year} {2017})}\BibitemShut {NoStop}%
\bibitem [{\citenamefont {Zhang}\ \emph
  {et~al.}(2020{\natexlab{a}})\citenamefont {Zhang}, \citenamefont {Yang},
  \citenamefont {Gong}, \citenamefont {Pan},\ and\ \citenamefont
  {Wang}}]{Zhang2020c}%
  \BibitemOpen
  \bibfield  {author} {\bibinfo {author} {\bibfnamefont {L.}~\bibnamefont
  {Zhang}}, \bibinfo {author} {\bibfnamefont {Z.}~\bibnamefont {Yang}},
  \bibinfo {author} {\bibfnamefont {T.}~\bibnamefont {Gong}}, \bibinfo {author}
  {\bibfnamefont {R.}~\bibnamefont {Pan}}, \ and\ \bibinfo {author}
  {\bibfnamefont {H.}~\bibnamefont {Wang}},\ }\href {\doibase
  10.1039/d0ta01999b} {\bibfield  {journal} {\bibinfo  {journal} {Journal of
  Materials Chemistry A}\ }\textbf {\bibinfo {volume} {8}},\ \bibinfo {pages}
  {8813} (\bibinfo {year} {2020}{\natexlab{a}})}\BibitemShut {NoStop}%
\bibitem [{\citenamefont {Din}\ \emph {et~al.}(2019)\citenamefont {Din},
  \citenamefont {Idrees}, \citenamefont {Albar}, \citenamefont {Shafiq},
  \citenamefont {Ahmad}, \citenamefont {Nguyen},\ and\ \citenamefont
  {Amin}}]{Din2019}%
  \BibitemOpen
  \bibfield  {author} {\bibinfo {author} {\bibfnamefont {H.~U.}\ \bibnamefont
  {Din}}, \bibinfo {author} {\bibfnamefont {M.}~\bibnamefont {Idrees}},
  \bibinfo {author} {\bibfnamefont {A.}~\bibnamefont {Albar}}, \bibinfo
  {author} {\bibfnamefont {M.}~\bibnamefont {Shafiq}}, \bibinfo {author}
  {\bibfnamefont {I.}~\bibnamefont {Ahmad}}, \bibinfo {author} {\bibfnamefont
  {C.~V.}\ \bibnamefont {Nguyen}}, \ and\ \bibinfo {author} {\bibfnamefont
  {B.}~\bibnamefont {Amin}},\ }\href {\doibase 10.1103/PhysRevB.100.165425}
  {\bibfield  {journal} {\bibinfo  {journal} {Physical Review B}\ }\textbf
  {\bibinfo {volume} {100}},\ \bibinfo {pages} {165425} (\bibinfo {year}
  {2019})}\BibitemShut {NoStop}%
\bibitem [{\citenamefont {Chen}\ \emph {et~al.}(2020)\citenamefont {Chen},
  \citenamefont {Wu}, \citenamefont {Ma}, \citenamefont {Hu},\ and\
  \citenamefont {Yang}}]{Chen2020}%
  \BibitemOpen
  \bibfield  {author} {\bibinfo {author} {\bibfnamefont {J.}~\bibnamefont
  {Chen}}, \bibinfo {author} {\bibfnamefont {K.}~\bibnamefont {Wu}}, \bibinfo
  {author} {\bibfnamefont {H.}~\bibnamefont {Ma}}, \bibinfo {author}
  {\bibfnamefont {W.}~\bibnamefont {Hu}}, \ and\ \bibinfo {author}
  {\bibfnamefont {J.}~\bibnamefont {Yang}},\ }\href {\doibase
  10.1039/d0ra00674b} {\bibfield  {journal} {\bibinfo  {journal} {RSC
  Advances}\ }\textbf {\bibinfo {volume} {10}},\ \bibinfo {pages} {6388}
  (\bibinfo {year} {2020})}\BibitemShut {NoStop}%
\bibitem [{\citenamefont {Manchon}\ \emph {et~al.}(2019)\citenamefont
  {Manchon}, \citenamefont {Zelezn{\'{y}}}, \citenamefont {Miron},
  \citenamefont {Jungwirth}, \citenamefont {Sinova}, \citenamefont {Thiaville},
  \citenamefont {Garello},\ and\ \citenamefont {Gambardella}}]{Manchon2019}%
  \BibitemOpen
  \bibfield  {author} {\bibinfo {author} {\bibfnamefont {A.}~\bibnamefont
  {Manchon}}, \bibinfo {author} {\bibfnamefont {J.}~\bibnamefont
  {Zelezn{\'{y}}}}, \bibinfo {author} {\bibfnamefont {M.}~\bibnamefont
  {Miron}}, \bibinfo {author} {\bibfnamefont {T.}~\bibnamefont {Jungwirth}},
  \bibinfo {author} {\bibfnamefont {J.}~\bibnamefont {Sinova}}, \bibinfo
  {author} {\bibfnamefont {A.}~\bibnamefont {Thiaville}}, \bibinfo {author}
  {\bibfnamefont {K.}~\bibnamefont {Garello}}, \ and\ \bibinfo {author}
  {\bibfnamefont {P.}~\bibnamefont {Gambardella}},\ }\href {\doibase
  10.1103/RevModPhys.91.035004} {\bibfield  {journal} {\bibinfo  {journal}
  {Review of Modern Physics}\ }\textbf {\bibinfo {volume} {91}},\ \bibinfo
  {pages} {035004} (\bibinfo {year} {2019})}\BibitemShut {NoStop}%
\bibitem [{\citenamefont {Smaili}\ \emph {et~al.}(2020)\citenamefont {Smaili},
  \citenamefont {Laref}, \citenamefont {Garcia}, \citenamefont
  {Schwingenschlogl}, \citenamefont {Roche},\ and\ \citenamefont
  {Manchon}}]{Smaili2020}%
  \BibitemOpen
  \bibfield  {author} {\bibinfo {author} {\bibfnamefont {I.}~\bibnamefont
  {Smaili}}, \bibinfo {author} {\bibfnamefont {S.}~\bibnamefont {Laref}},
  \bibinfo {author} {\bibfnamefont {J.~H.}\ \bibnamefont {Garcia}}, \bibinfo
  {author} {\bibfnamefont {U.}~\bibnamefont {Schwingenschlogl}}, \bibinfo
  {author} {\bibfnamefont {S.}~\bibnamefont {Roche}}, \ and\ \bibinfo {author}
  {\bibfnamefont {A.}~\bibnamefont {Manchon}},\ }\href
  {http://arxiv.org/abs/2007.07579} {\bibfield  {journal} {\bibinfo  {journal}
  {arXiv preprint arXiV:2007.07579v1}\ ,\ \bibinfo {pages} {1}} (\bibinfo
  {year} {2020})},\ \Eprint {http://arxiv.org/abs/2007.07579}
  {arXiv:2007.07579} \BibitemShut {NoStop}%
\bibitem [{\citenamefont {Liang}\ \emph {et~al.}(2020)\citenamefont {Liang},
  \citenamefont {Wang}, \citenamefont {Du}, \citenamefont {Hallal},
  \citenamefont {Garcia}, \citenamefont {Chshiev}, \citenamefont {Fert},\ and\
  \citenamefont {Yang}}]{Liang2020}%
  \BibitemOpen
  \bibfield  {author} {\bibinfo {author} {\bibfnamefont {J.}~\bibnamefont
  {Liang}}, \bibinfo {author} {\bibfnamefont {W.}~\bibnamefont {Wang}},
  \bibinfo {author} {\bibfnamefont {H.}~\bibnamefont {Du}}, \bibinfo {author}
  {\bibfnamefont {A.}~\bibnamefont {Hallal}}, \bibinfo {author} {\bibfnamefont
  {K.}~\bibnamefont {Garcia}}, \bibinfo {author} {\bibfnamefont
  {M.}~\bibnamefont {Chshiev}}, \bibinfo {author} {\bibfnamefont
  {A.}~\bibnamefont {Fert}}, \ and\ \bibinfo {author} {\bibfnamefont
  {H.}~\bibnamefont {Yang}},\ }\href {\doibase 10.1103/PhysRevB.101.184401}
  {\bibfield  {journal} {\bibinfo  {journal} {Physical Review B}\ }\textbf
  {\bibinfo {volume} {101}},\ \bibinfo {pages} {184401} (\bibinfo {year}
  {2020})}\BibitemShut {NoStop}%
\bibitem [{\citenamefont {Zhang}\ \emph
  {et~al.}(2020{\natexlab{b}})\citenamefont {Zhang}, \citenamefont {Zhang},
  \citenamefont {Mi},\ and\ \citenamefont {Wang}}]{Zhang2020d}%
  \BibitemOpen
  \bibfield  {author} {\bibinfo {author} {\bibfnamefont {F.}~\bibnamefont
  {Zhang}}, \bibinfo {author} {\bibfnamefont {H.}~\bibnamefont {Zhang}},
  \bibinfo {author} {\bibfnamefont {W.}~\bibnamefont {Mi}}, \ and\ \bibinfo
  {author} {\bibfnamefont {X.}~\bibnamefont {Wang}},\ }\href {\doibase
  10.1039/d0cp00174k} {\bibfield  {journal} {\bibinfo  {journal} {Phys. Chem.
  Chem. Phys.}\ }\textbf {\bibinfo {volume} {22}},\ \bibinfo {pages} {8647}
  (\bibinfo {year} {2020}{\natexlab{b}})}\BibitemShut {NoStop}%
\bibitem [{\citenamefont {Yuan}\ \emph {et~al.}(2020)\citenamefont {Yuan},
  \citenamefont {Yang}, \citenamefont {Cai}, \citenamefont {Wu}, \citenamefont
  {Chen}, \citenamefont {Yan},\ and\ \citenamefont {Shen}}]{Yuan2020c}%
  \BibitemOpen
  \bibfield  {author} {\bibinfo {author} {\bibfnamefont {J.}~\bibnamefont
  {Yuan}}, \bibinfo {author} {\bibfnamefont {Y.}~\bibnamefont {Yang}}, \bibinfo
  {author} {\bibfnamefont {Y.}~\bibnamefont {Cai}}, \bibinfo {author}
  {\bibfnamefont {Y.}~\bibnamefont {Wu}}, \bibinfo {author} {\bibfnamefont
  {Y.}~\bibnamefont {Chen}}, \bibinfo {author} {\bibfnamefont {X.}~\bibnamefont
  {Yan}}, \ and\ \bibinfo {author} {\bibfnamefont {L.}~\bibnamefont {Shen}},\
  }\href {\doibase 10.1103/PhysRevB.101.094420} {\bibfield  {journal} {\bibinfo
   {journal} {Physical Review B}\ }\textbf {\bibinfo {volume} {101}},\ \bibinfo
  {pages} {094420} (\bibinfo {year} {2020})}\BibitemShut {NoStop}%
\bibitem [{\citenamefont {Hamann}(1979)}]{hamann1979semiconductor}%
  \BibitemOpen
  \bibfield  {author} {\bibinfo {author} {\bibfnamefont {D.}~\bibnamefont
  {Hamann}},\ }\href {\doibase https://doi.org/10.1103/PhysRevLett.42.662}
  {\bibfield  {journal} {\bibinfo  {journal} {Physical Review Letters}\
  }\textbf {\bibinfo {volume} {42}},\ \bibinfo {pages} {662} (\bibinfo {year}
  {1979})}\BibitemShut {NoStop}%
\bibitem [{\citenamefont {Wimmer}\ \emph {et~al.}(1981)\citenamefont {Wimmer},
  \citenamefont {Krakauer}, \citenamefont {Weinert},\ and\ \citenamefont
  {Freeman}}]{wimmer1981full}%
  \BibitemOpen
  \bibfield  {author} {\bibinfo {author} {\bibfnamefont {E.}~\bibnamefont
  {Wimmer}}, \bibinfo {author} {\bibfnamefont {H.}~\bibnamefont {Krakauer}},
  \bibinfo {author} {\bibfnamefont {M.}~\bibnamefont {Weinert}}, \ and\
  \bibinfo {author} {\bibfnamefont {A.~J.}\ \bibnamefont {Freeman}},\ }\href
  {\doibase https://doi.org/10.1103/PhysRevB.24.864} {\bibfield  {journal}
  {\bibinfo  {journal} {Physical Review B}\ }\textbf {\bibinfo {volume} {24}},\
  \bibinfo {pages} {864} (\bibinfo {year} {1981})}\BibitemShut {NoStop}%
\bibitem [{fla()}]{flapw}%
  \BibitemOpen
  \href {https://www.flapw.de} {\enquote {\bibinfo {title} {Fleur},}\
  }\BibitemShut {NoStop}%
\bibitem [{\citenamefont {Ma}\ \emph {et~al.}(2012)\citenamefont {Ma},
  \citenamefont {Dai}, \citenamefont {Guo}, \citenamefont {Niu}, \citenamefont
  {Zhu},\ and\ \citenamefont {Huang}}]{Ma2012evidence}%
  \BibitemOpen
  \bibfield  {author} {\bibinfo {author} {\bibfnamefont {Y.}~\bibnamefont
  {Ma}}, \bibinfo {author} {\bibfnamefont {Y.}~\bibnamefont {Dai}}, \bibinfo
  {author} {\bibfnamefont {M.}~\bibnamefont {Guo}}, \bibinfo {author}
  {\bibfnamefont {C.}~\bibnamefont {Niu}}, \bibinfo {author} {\bibfnamefont
  {Y.}~\bibnamefont {Zhu}}, \ and\ \bibinfo {author} {\bibfnamefont
  {B.}~\bibnamefont {Huang}},\ }\href {\doibase
  https://doi.org/10.1021/nn204667z} {\bibfield  {journal} {\bibinfo  {journal}
  {ACS Nano}\ }\textbf {\bibinfo {volume} {6}},\ \bibinfo {pages} {1695}
  (\bibinfo {year} {2012})}\BibitemShut {NoStop}%
\bibitem [{\citenamefont {Feng}\ \emph {et~al.}(2012)\citenamefont {Feng},
  \citenamefont {Peng}, \citenamefont {Wu}, \citenamefont {Sun}, \citenamefont
  {Hu}, \citenamefont {Lin}, \citenamefont {Dai}, \citenamefont {Yang},\ and\
  \citenamefont {Xie}}]{Feng2012giant}%
  \BibitemOpen
  \bibfield  {author} {\bibinfo {author} {\bibfnamefont {J.}~\bibnamefont
  {Feng}}, \bibinfo {author} {\bibfnamefont {L.}~\bibnamefont {Peng}}, \bibinfo
  {author} {\bibfnamefont {C.}~\bibnamefont {Wu}}, \bibinfo {author}
  {\bibfnamefont {X.}~\bibnamefont {Sun}}, \bibinfo {author} {\bibfnamefont
  {S.}~\bibnamefont {Hu}}, \bibinfo {author} {\bibfnamefont {C.}~\bibnamefont
  {Lin}}, \bibinfo {author} {\bibfnamefont {J.}~\bibnamefont {Dai}}, \bibinfo
  {author} {\bibfnamefont {J.}~\bibnamefont {Yang}}, \ and\ \bibinfo {author}
  {\bibfnamefont {Y.}~\bibnamefont {Xie}},\ }\href {\doibase
  https://doi.org/10.1002/adma.201104681} {\bibfield  {journal} {\bibinfo
  {journal} {Advanced Materials}\ }\textbf {\bibinfo {volume} {24}},\ \bibinfo
  {pages} {1969} (\bibinfo {year} {2012})}\BibitemShut {NoStop}%
\bibitem [{\citenamefont {Zhang}\ \emph {et~al.}(2013)\citenamefont {Zhang},
  \citenamefont {Liu},\ and\ \citenamefont {Lau}}]{Zhang2013dimension}%
  \BibitemOpen
  \bibfield  {author} {\bibinfo {author} {\bibfnamefont {H.}~\bibnamefont
  {Zhang}}, \bibinfo {author} {\bibfnamefont {L.-M.}\ \bibnamefont {Liu}}, \
  and\ \bibinfo {author} {\bibfnamefont {W.-M.}\ \bibnamefont {Lau}},\ }\href
  {\doibase https://doi.org/10.1039/C3TA12098H} {\bibfield  {journal} {\bibinfo
   {journal} {Journal of Materials Chemistry A}\ }\textbf {\bibinfo {volume}
  {1}},\ \bibinfo {pages} {10821} (\bibinfo {year} {2013})}\BibitemShut
  {NoStop}%
\bibitem [{\citenamefont {Pan}(2014)}]{Pan2014electronic}%
  \BibitemOpen
  \bibfield  {author} {\bibinfo {author} {\bibfnamefont {H.}~\bibnamefont
  {Pan}},\ }\href {\doibase https://doi.org/10.1021/jp503030b} {\bibfield
  {journal} {\bibinfo  {journal} {The Journal of Physical Chemistry C}\
  }\textbf {\bibinfo {volume} {118}},\ \bibinfo {pages} {13248} (\bibinfo
  {year} {2014})}\BibitemShut {NoStop}%
\bibitem [{\citenamefont {Oswald}\ \emph {et~al.}(1985)\citenamefont {Oswald},
  \citenamefont {Zeller}, \citenamefont {Braspenning},\ and\ \citenamefont
  {Dederichs}}]{oswald1985interaction}%
  \BibitemOpen
  \bibfield  {author} {\bibinfo {author} {\bibfnamefont {A.}~\bibnamefont
  {Oswald}}, \bibinfo {author} {\bibfnamefont {R.}~\bibnamefont {Zeller}},
  \bibinfo {author} {\bibfnamefont {P.}~\bibnamefont {Braspenning}}, \ and\
  \bibinfo {author} {\bibfnamefont {P.}~\bibnamefont {Dederichs}},\ }\href
  {https://iopscience.iop.org/article/10.1088/0305-4608/15/1/021/meta}
  {\bibfield  {journal} {\bibinfo  {journal} {Journal of Physics F: Metal
  Physics}\ }\textbf {\bibinfo {volume} {15}},\ \bibinfo {pages} {193}
  (\bibinfo {year} {1985})}\BibitemShut {NoStop}%
\bibitem [{\citenamefont {Liechtenstein}\ \emph {et~al.}(1987)\citenamefont
  {Liechtenstein}, \citenamefont {Katsnelson}, \citenamefont {Antropov},\ and\
  \citenamefont {Gubanov}}]{liechtenstein1987local}%
  \BibitemOpen
  \bibfield  {author} {\bibinfo {author} {\bibfnamefont {A.~I.}\ \bibnamefont
  {Liechtenstein}}, \bibinfo {author} {\bibfnamefont {M.}~\bibnamefont
  {Katsnelson}}, \bibinfo {author} {\bibfnamefont {V.}~\bibnamefont
  {Antropov}}, \ and\ \bibinfo {author} {\bibfnamefont {V.}~\bibnamefont
  {Gubanov}},\ }\href {\doibase https://doi.org/10.1016/0304-8853(87)90721-9}
  {\bibfield  {journal} {\bibinfo  {journal} {Journal of Magnetism and Magnetic
  Materials}\ }\textbf {\bibinfo {volume} {67}},\ \bibinfo {pages} {65}
  (\bibinfo {year} {1987})}\BibitemShut {NoStop}%
\bibitem [{\citenamefont {Li}\ \emph {et~al.}(1990)\citenamefont {Li},
  \citenamefont {Freeman}, \citenamefont {Jansen},\ and\ \citenamefont
  {Fu}}]{li1990magnetic}%
  \BibitemOpen
  \bibfield  {author} {\bibinfo {author} {\bibfnamefont {C.}~\bibnamefont
  {Li}}, \bibinfo {author} {\bibfnamefont {A.~J.}\ \bibnamefont {Freeman}},
  \bibinfo {author} {\bibfnamefont {H.}~\bibnamefont {Jansen}}, \ and\ \bibinfo
  {author} {\bibfnamefont {C.}~\bibnamefont {Fu}},\ }\href {\doibase
  https://doi.org/10.1103/PhysRevB.42.5433} {\bibfield  {journal} {\bibinfo
  {journal} {Physical Review B}\ }\textbf {\bibinfo {volume} {42}},\ \bibinfo
  {pages} {5433} (\bibinfo {year} {1990})}\BibitemShut {NoStop}%
\bibitem [{\citenamefont {Sandratskii}(1991)}]{sandratskii1991symmetry}%
  \BibitemOpen
  \bibfield  {author} {\bibinfo {author} {\bibfnamefont {L.}~\bibnamefont
  {Sandratskii}},\ }\href
  {https://iopscience.iop.org/article/10.1088/0953-8984/3/44/004} {\bibfield
  {journal} {\bibinfo  {journal} {Journal of Physics: Condensed Matter}\
  }\textbf {\bibinfo {volume} {3}},\ \bibinfo {pages} {8565} (\bibinfo {year}
  {1991})}\BibitemShut {NoStop}%
\bibitem [{\citenamefont {Kurz}\ \emph {et~al.}(2004)\citenamefont {Kurz},
  \citenamefont {F{\"{o}}rster}, \citenamefont {Nordstr{\"{o}}m}, \citenamefont
  {Bihlmayer},\ and\ \citenamefont {Bl{\"{u}}gel}}]{Kurz2004}%
  \BibitemOpen
  \bibfield  {author} {\bibinfo {author} {\bibfnamefont {P.}~\bibnamefont
  {Kurz}}, \bibinfo {author} {\bibfnamefont {F.}~\bibnamefont {F{\"{o}}rster}},
  \bibinfo {author} {\bibfnamefont {L.}~\bibnamefont {Nordstr{\"{o}}m}},
  \bibinfo {author} {\bibfnamefont {G.}~\bibnamefont {Bihlmayer}}, \ and\
  \bibinfo {author} {\bibfnamefont {S.}~\bibnamefont {Bl{\"{u}}gel}},\ }\href
  {\doibase 10.1103/PhysRevB.69.024415} {\bibfield  {journal} {\bibinfo
  {journal} {Physical Review B}\ }\textbf {\bibinfo {volume} {69}},\ \bibinfo
  {pages} {024415} (\bibinfo {year} {2004})}\BibitemShut {NoStop}%
\bibitem [{\citenamefont {Heide}\ \emph {et~al.}(2009)\citenamefont {Heide},
  \citenamefont {Bihlmayer},\ and\ \citenamefont {Bl{\"{u}}gel}}]{Heide2009}%
  \BibitemOpen
  \bibfield  {author} {\bibinfo {author} {\bibfnamefont {M.}~\bibnamefont
  {Heide}}, \bibinfo {author} {\bibfnamefont {G.}~\bibnamefont {Bihlmayer}}, \
  and\ \bibinfo {author} {\bibfnamefont {S.}~\bibnamefont {Bl{\"{u}}gel}},\
  }\href {\doibase 10.1016/j.physb.2009.06.070} {\bibfield  {journal} {\bibinfo
   {journal} {Physica B: Condensed Matter}\ }\textbf {\bibinfo {volume}
  {404}},\ \bibinfo {pages} {2678} (\bibinfo {year} {2009})}\BibitemShut
  {NoStop}%
\bibitem [{\citenamefont {Zimmermann}\ \emph {et~al.}(2014)\citenamefont
  {Zimmermann}, \citenamefont {Heide}, \citenamefont {Bihlmayer},\ and\
  \citenamefont {Bl{\"u}gel}}]{zimmermann2014first}%
  \BibitemOpen
  \bibfield  {author} {\bibinfo {author} {\bibfnamefont {B.}~\bibnamefont
  {Zimmermann}}, \bibinfo {author} {\bibfnamefont {M.}~\bibnamefont {Heide}},
  \bibinfo {author} {\bibfnamefont {G.}~\bibnamefont {Bihlmayer}}, \ and\
  \bibinfo {author} {\bibfnamefont {S.}~\bibnamefont {Bl{\"u}gel}},\ }\href
  {\doibase https://doi.org/10.1103/PhysRevB.90.115427} {\bibfield  {journal}
  {\bibinfo  {journal} {Physical Review B}\ }\textbf {\bibinfo {volume} {90}},\
  \bibinfo {pages} {115427} (\bibinfo {year} {2014})}\BibitemShut {NoStop}%
\bibitem [{\citenamefont {Belabbes}\ \emph {et~al.}(2016)\citenamefont
  {Belabbes}, \citenamefont {Bihlmayer}, \citenamefont {Bechstedt},
  \citenamefont {Bl{\"{u}}gel},\ and\ \citenamefont {Manchon}}]{Belabbes2016}%
  \BibitemOpen
  \bibfield  {author} {\bibinfo {author} {\bibfnamefont {A.}~\bibnamefont
  {Belabbes}}, \bibinfo {author} {\bibfnamefont {G.}~\bibnamefont {Bihlmayer}},
  \bibinfo {author} {\bibfnamefont {F.}~\bibnamefont {Bechstedt}}, \bibinfo
  {author} {\bibfnamefont {S.}~\bibnamefont {Bl{\"{u}}gel}}, \ and\ \bibinfo
  {author} {\bibfnamefont {A.}~\bibnamefont {Manchon}},\ }\href {\doibase
  10.1103/PhysRevLett.117.247202} {\bibfield  {journal} {\bibinfo  {journal}
  {Physical Review Letters}\ }\textbf {\bibinfo {volume} {117}},\ \bibinfo
  {pages} {247202} (\bibinfo {year} {2016})}\BibitemShut {NoStop}%
\bibitem [{\citenamefont {Evans}\ \emph {et~al.}(2014)\citenamefont {Evans},
  \citenamefont {Fan}, \citenamefont {Chureemart}, \citenamefont {Ostler},
  \citenamefont {Ellis},\ and\ \citenamefont {Chantrell}}]{evans2014}%
  \BibitemOpen
  \bibfield  {author} {\bibinfo {author} {\bibfnamefont {R.~F.~L.}\
  \bibnamefont {Evans}}, \bibinfo {author} {\bibfnamefont {W.~J.}\ \bibnamefont
  {Fan}}, \bibinfo {author} {\bibfnamefont {P.}~\bibnamefont {Chureemart}},
  \bibinfo {author} {\bibfnamefont {T.~A.}\ \bibnamefont {Ostler}}, \bibinfo
  {author} {\bibfnamefont {M.~O.~A.}\ \bibnamefont {Ellis}}, \ and\ \bibinfo
  {author} {\bibfnamefont {R.~W.}\ \bibnamefont {Chantrell}},\ }\href {\doibase
  10.1088/0953-8984/26/10/103202} {\bibfield  {journal} {\bibinfo  {journal}
  {Journal of Physics: Condensed Matter}\ }\textbf {\bibinfo {volume} {26}},\
  \bibinfo {pages} {103202} (\bibinfo {year} {2014})}\BibitemShut {NoStop}%
\bibitem [{\citenamefont {Berg}\ and\ \citenamefont
  {Luscher}(1981)}]{berg1981}%
  \BibitemOpen
  \bibfield  {author} {\bibinfo {author} {\bibfnamefont {B.}~\bibnamefont
  {Berg}}\ and\ \bibinfo {author} {\bibfnamefont {M.}~\bibnamefont {Luscher}},\
  }\href@noop {} {\bibfield  {journal} {\bibinfo  {journal} {Nuclear Physics
  B}\ }\textbf {\bibinfo {volume} {190}},\ \bibinfo {pages} {412} (\bibinfo
  {year} {1981})}\BibitemShut {NoStop}%
\bibitem [{\citenamefont {B{\"{o}}ttcher}\ \emph {et~al.}(2018)\citenamefont
  {B{\"{o}}ttcher}, \citenamefont {Heinze}, \citenamefont {Egorov},\ and\
  \citenamefont {Sinova}}]{bottcher2018}%
  \BibitemOpen
  \bibfield  {author} {\bibinfo {author} {\bibfnamefont {M.}~\bibnamefont
  {B{\"{o}}ttcher}}, \bibinfo {author} {\bibfnamefont {S.}~\bibnamefont
  {Heinze}}, \bibinfo {author} {\bibfnamefont {S.}~\bibnamefont {Egorov}}, \
  and\ \bibinfo {author} {\bibfnamefont {J.}~\bibnamefont {Sinova}},\
  }\href@noop {} {\bibfield  {journal} {\bibinfo  {journal} {New Journal of
  Physics}\ }\textbf {\bibinfo {volume} {20}},\ \bibinfo {pages} {103014}
  (\bibinfo {year} {2018})}\BibitemShut {NoStop}%
\bibitem [{\citenamefont {M{\"{u}}ller}\ \emph {et~al.}(2019)\citenamefont
  {M{\"{u}}ller}, \citenamefont {Hoffmann}, \citenamefont {Di{\ss}elkamp},
  \citenamefont {Sch{\"{u}}rhoff}, \citenamefont {Mavros}, \citenamefont
  {Sallermann}, \citenamefont {Kiselev}, \citenamefont {J{\'{o}}nsson},\ and\
  \citenamefont {Bl{\"{u}}gel}}]{muller2019}%
  \BibitemOpen
  \bibfield  {author} {\bibinfo {author} {\bibfnamefont {G.~P.}\ \bibnamefont
  {M{\"{u}}ller}}, \bibinfo {author} {\bibfnamefont {M.}~\bibnamefont
  {Hoffmann}}, \bibinfo {author} {\bibfnamefont {C.}~\bibnamefont
  {Di{\ss}elkamp}}, \bibinfo {author} {\bibfnamefont {D.}~\bibnamefont
  {Sch{\"{u}}rhoff}}, \bibinfo {author} {\bibfnamefont {S.}~\bibnamefont
  {Mavros}}, \bibinfo {author} {\bibfnamefont {M.}~\bibnamefont {Sallermann}},
  \bibinfo {author} {\bibfnamefont {N.~S.}\ \bibnamefont {Kiselev}}, \bibinfo
  {author} {\bibfnamefont {H.}~\bibnamefont {J{\'{o}}nsson}}, \ and\ \bibinfo
  {author} {\bibfnamefont {S.}~\bibnamefont {Bl{\"{u}}gel}},\ }\href {\doibase
  10.1103/PhysRevB.99.224414} {\bibfield  {journal} {\bibinfo  {journal}
  {Physical Review B}\ }\textbf {\bibinfo {volume} {99}},\ \bibinfo {pages}
  {224414} (\bibinfo {year} {2019})}\BibitemShut {NoStop}%
\bibitem [{\citenamefont {Shen}\ \emph {et~al.}(2020)\citenamefont {Shen},
  \citenamefont {Li}, \citenamefont {Xia},\ and\ \citenamefont
  {Qiu}}]{lshen2020}%
  \BibitemOpen
  \bibfield  {author} {\bibinfo {author} {\bibfnamefont {L.}~\bibnamefont
  {Shen}}, \bibinfo {author} {\bibfnamefont {X.}~\bibnamefont {Li}}, \bibinfo
  {author} {\bibfnamefont {J.}~\bibnamefont {Xia}}, \ and\ \bibinfo {author}
  {\bibfnamefont {L.}~\bibnamefont {Qiu}},\ }\href@noop {} {\bibfield
  {journal} {\bibinfo  {journal} {arXiv preprint arXiv:2005.11924}\ } (\bibinfo
  {year} {2020})},\ \Eprint {http://arxiv.org/abs/arXiv:2005.11924v2}
  {arXiv:arXiv:2005.11924v2} \BibitemShut {NoStop}%
\bibitem [{\citenamefont {Rozsa}\ \emph {et~al.}(2016)\citenamefont {Rozsa},
  \citenamefont {Simon}, \citenamefont {Palotas}, \citenamefont {Udvardi},\
  and\ \citenamefont {Szunyogh}}]{rozsa2016}%
  \BibitemOpen
  \bibfield  {author} {\bibinfo {author} {\bibfnamefont {L.}~\bibnamefont
  {Rozsa}}, \bibinfo {author} {\bibfnamefont {E.}~\bibnamefont {Simon}},
  \bibinfo {author} {\bibfnamefont {K.}~\bibnamefont {Palotas}}, \bibinfo
  {author} {\bibfnamefont {L.}~\bibnamefont {Udvardi}}, \ and\ \bibinfo
  {author} {\bibfnamefont {L.}~\bibnamefont {Szunyogh}},\ }\href {\doibase
  10.1103/PhysRevB.93.024417} {\bibfield  {journal} {\bibinfo  {journal}
  {Physical Review B}\ }\textbf {\bibinfo {volume} {93}},\ \bibinfo {pages}
  {024417} (\bibinfo {year} {2016})}\BibitemShut {NoStop}%
\bibitem [{\citenamefont {Grigoriev}\ \emph {et~al.}(2011)\citenamefont
  {Grigoriev}, \citenamefont {Moskvin}, \citenamefont {Dyadkin}, \citenamefont
  {Lamago}, \citenamefont {Wolf}, \citenamefont {Eckerlebe},\ and\
  \citenamefont {Maleyev}}]{grigoriev2011}%
  \BibitemOpen
  \bibfield  {author} {\bibinfo {author} {\bibfnamefont {S.~V.}\ \bibnamefont
  {Grigoriev}}, \bibinfo {author} {\bibfnamefont {E.~V.}\ \bibnamefont
  {Moskvin}}, \bibinfo {author} {\bibfnamefont {V.~A.}\ \bibnamefont
  {Dyadkin}}, \bibinfo {author} {\bibfnamefont {D.}~\bibnamefont {Lamago}},
  \bibinfo {author} {\bibfnamefont {T.}~\bibnamefont {Wolf}}, \bibinfo {author}
  {\bibfnamefont {H.}~\bibnamefont {Eckerlebe}}, \ and\ \bibinfo {author}
  {\bibfnamefont {S.~V.}\ \bibnamefont {Maleyev}},\ }\href {\doibase
  10.1103/PhysRevB.83.224411} {\bibfield  {journal} {\bibinfo  {journal}
  {Physical Review B}\ }\textbf {\bibinfo {volume} {83}},\ \bibinfo {pages}
  {224411} (\bibinfo {year} {2011})}\BibitemShut {NoStop}%
\bibitem [{\citenamefont {Takei}\ and\ \citenamefont
  {Tserkovnyak}(2014)}]{Takei2014}%
  \BibitemOpen
  \bibfield  {author} {\bibinfo {author} {\bibfnamefont {S.}~\bibnamefont
  {Takei}}\ and\ \bibinfo {author} {\bibfnamefont {Y.}~\bibnamefont
  {Tserkovnyak}},\ }\href {\doibase 10.1103/PhysRevLett.112.227201} {\bibfield
  {journal} {\bibinfo  {journal} {Physical Review Letters}\ }\textbf {\bibinfo
  {volume} {112}},\ \bibinfo {pages} {227201} (\bibinfo {year}
  {2014})}\BibitemShut {NoStop}%
\end{thebibliography}%
\clearpage
\end{document}